\documentclass{article}

\usepackage{microtype}
\usepackage{graphicx}
\usepackage{subfigure}
\usepackage{booktabs} 
\usepackage{amsfonts,amssymb,amsthm,amsmath}
\usepackage{nicefrac}  
\usepackage{natbib}
\usepackage{hyperref}
\usepackage{tikz}

\usepackage{fullpage}
\usepackage{xspace}
\usepackage{amsthm}
\usepackage{lineno}

\usepackage{algorithm}
\usepackage{algorithmic}

\providecommand{\keywords}[1]
{
  \small	
  \textbf{\textit{Keywords---}} #1
}

\usepackage{authblk}

\title{Exploring the non-stationarity of coastal sea level probability distributions}

\date{}
\begin{document}

\author[1]{Fabrizio Falasca \thanks{Corresponding Author: fabri.falasca@nyu.edu}}
\author[1]{Andrew Brettin}
\author[1]{Laure Zanna}
\author[2,3]{Stephen M. Griffies}
\author[4]{Jianjun Yin}
\author[2]{Ming Zhao}

\affil[1]{Courant Institute of Mathematical Sciences, New York University, New York, NY,  USA}
\affil[2]{NOAA Geophysical Fluid Dynamics Laboratory, Princeton, New Jersey, USA}
\affil[3]{Princeton University Atmospheric and Oceanic Sciences Program, Princeton University, Princeton, New Jersey, USA}
\affil[4]{Department of Geosciences, The University of Arizona, Tucson, Arizona, USA}

\maketitle

\abstract{Studies agree on a significant global mean sea level rise in the 20th century and its recent 21st century acceleration in the satellite record. At regional scale, the evolution of sea level probability distributions is often assumed to be dominated by changes in the mean. However, a quantification of changes in distributional shapes in a changing climate is currently missing. To this end, we propose a novel framework quantifying significant changes in probability distributions from time series data. The framework first quantifies linear trends in quantiles through quantile regression. Quantile slopes are then projected onto a set of four \textit{orthogonal} polynomials quantifying how such changes can be explained by \textit{independent} shifts in the first four statistical moments. The framework proposed is theoretically founded, general and can be applied to any climate observable with close-to-linear changes in distributions. We focus on observations and a coupled climate model (GFDL-CM4). In the historical period, trends in coastal daily sea level have been driven mainly by changes in the mean and can therefore be explained by a shift of the distribution with no change in shape. In the modeled world, robust changes in higher order moments emerge with increasing $\text{CO}_{2}$ concentration. Such changes are driven in part by ocean circulation alone and get amplified by sea level pressure fluctuations, with  possible consequences for sea level extremes attribution studies.}

\keywords{Non-stationary processes; non-Gaussianity; probability distributions; sea level; climate models}

\section{Introduction} 

Regional and global sea level are  affected by both natural climate variations and anthropogenic climate change, with possible repercussions on densely populated coastal communities of great concern. Anthropogenic global warming recently motivated a number of studies characterizing rates and causes of sea level rise since the early 20th century. A reconstruction of global sea level trends since 1900 has been recently proposed by \cite{Dangendorf2019} and \cite{frederikse2020causes}. Both studies show a robust increase in mean sea level with trends of $1.6 \pm 0.4$ mm/yr and $1.56 \pm 0.33$ mm/yr over the periods 1900-2015 and 1900-2018, respectively. Such global trends are not constant over time and marked by an acceleration over the recent three decades, as quantified by the $3.1 \pm 0.3$ mm/yr trend measured by altimetry since 1993 \citep{WCRP}. This acceleration has been mainly ascribed to an increase in ocean heat uptake driven by changes in Southern Hemisphere westerlies, as well as increased mass loss over Greenland \citep{ipccChapter9,Dangendorf2019,frederikse2020causes}. \\

Globally, the causes of trends in sea level since 1900 are reasonably well understood, with the largest contribution coming from glaciers ($52\%$), followed by ocean thermal expansion ($32\%$) and Greenland ice sheet mass loss ($29\%$) plus a negative contribution coming from the land water storage \citep{frederikse2020causes}. Regionally, on the other hand, sea level can be strongly affected by local variability in patterns of ocean circulation and water masses such as those caused by fluctuations in winds, ocean heating and moisture fluxes \citep{Hamlington,Han}. Such dynamical changes can potentially mask sea level trends in regions dominated by large multidecadal variability. A known example is the observed large trend in sea level in the western Pacific ocean since 1990, mainly reflecting phases of the Pacific Decadal Oscillation (PDO) \citep{Han,Zhang,Merrifield}. Changes not linked to ocean and atmosphere dynamics also play a role. An example is the vertical land movement in some areas as a consequence of glacial isostatic adjustment (GIA) since the last ice age, causing relatively large differences in long term trends across basins \citep{Wang,Caron,Tamisiea}. Notwithstanding the large number of possible contributions, the sea level budget at tide gauges since 1958 has been recently closed in \cite{Wang}. The authors identified sterodynamic sea-level change as the main contributors to sea level rise in many locations, and GIA in few others \citep{Wang}
(as detailed in \citep{gregory}, sterodynamic changes arise from changes in ocean density and ocean circulation).\\

Trends in sea level at long time-scales also impact changes in weather-like extremes. Recently, many studies have focused on quantification of current and future changes in sea level extremes driven by storms, tides and waves \citep{Wahl,Buchanan,Rasmussen, Sweet,Tebaldi}. Such studies aim to characterize tails of sea level distributions in recent periods or in future scenarios via Extreme Value Theory (EVT) \citep{Coles}, and often quantify changes in extremes solely as a function of changes in distributional mean \citep{Tebaldi}. A shift in the mean sea level is in fact recognized to be the primary driver of changes in tails of the distributions \citep{Vousdoukas,Sreeraj2022}. Few studies also explored extreme value analysis by considering changes in the median and width of the fitted generalized extreme value distributions. Examples range from the work of \cite{wong2022} where the authors explored non-stationarity of extreme value statistics covarying with different climate variables, to \cite{Lee} quantifying links between frequencies of sea level extremes and global mean temperature. Similarly, \cite{Grinsted} investigated relationships between changes in extreme value analysis in storm surges and different predictors, from the PDO pattern to Quasi Biennal Oscillation among others.\\

While recent studies have focused on trends in the mean sea level or on large extremes through EVT, there has been less work quantifying how shapes of sea level probability distributions have been changing in the observational record and how they may change in a warming climate. 
This is no easy task, as quantifying trends in distributions and their significance under internal climate variability is not a well defined problem and many measures could be adopted. \cite{Cozannet} studied changes in sea level distributions in climate models projections focusing on changes in the cumulative distribution functions. Another option would be to directly estimate changes in moments. A useful, comprehensive methodology was recently proposed by \cite{mckinnon2016changing} where the authors proposed a unified framework to study both linear changes in quantiles and moments of summer temperature time series. Changes in quantiles can be quantified through the quantile regression approach used by \cite{Koenker,koenker2001quantile}. Temporal changes in  quantiles of the distributions can be further summarized through projections onto few polynomials linking changes in quantiles to changes in statistical moments. \cite{mckinnon2016changing} empirically showed that Legendre polynomials may be suitable functions for this purpose.\\

Motivated by quantifying changes in sea level rise and inspired by \cite{mckinnon2016changing}, here we propose an alternative/complementary route to study temporal changes in probability distributions by building upon \cite{mckinnon2016changing} in two ways. First, based on the work of \cite{Cornish}, we provide a general, analytical relationship between time changes in quantiles and the first four moments of a distribution. This analytical relationship allows us to build a theoretically founded methodology to explore changes in distributions. We show how the framework allows us to diagnose time changes in distributions as sums of \textit{independent} shifts in the first four statistical moments. Second, we study the significance of such changes in the presence of internal climate variability by accounting for multiple testings \citep{Benjamini}. This approach limits the number of false positives in the inference step to obtain trustworthy results even in the presence of a large number of significance tests \citep{Witten}. Importantly, such linear framework is shown to work well even in case of weak nonlinearities, with few examples shown later with synthetic data in Section \ref{subsec:synthetic} and in Appendix \ref{app:balboa_record} and \ref{app:Med-sea}.\\

We apply the proposed framework to daily tide gauges data in the 1970-2017 period, for which many locations have reliable data, and in a few locations with longer records. We therefore quantify changes in sea level distributions and assess their significance under the internal variability of the system.\\

Observational results are then complemented by output from the GFDL-CM4 climate model \citep{Held}. We first focus on the historical experiment and compare it with observations. We then investigate a transient experiment with 1$\%$ $\text{CO}_{2}$ increase per year, starting from a preindustrial global $\text{CO}_{2}$ concentration and extending up to quadrupling after 140 years. Crucially, the modeled trends can be further decomposed into different sea level components. We therefore explore sea level rise  as a result of ocean circulation only and when fluctuations in the atmospheric pressure at the sea surface (i.e., sea level pressure) are included.\\

The paper is organized as follows. In Section \ref{sec:data}, we present the data and the sea level decomposition adopted in this study. Section \ref{sec:methodology} presents the framework to study shifts in distributions. Results are presented in Section \ref{sec:results}, and  Section \ref{sec:conclusions_discussions} concludes the paper.

\section{Datasets} \label{sec:data}

\subsection{Observational tide gauge record}

We focus on local observations of daily sea level from tide gauge data provided by the University of Hawaii Sea Level Center \citep{Caldwell} (\url{https://uhslc.soest.hawaii.edu/}). We first consider the period 1970-2017 and keep only tide gauges with less than 20$\%$ missing values. Data gaps are not filled. This step reduces the number of time series considered from 116 to 94. The time range 1970-2017 was chosen as a compromise between retaining high-quality, daily sea level observations and sampling a large portion of coastal areas (e.g., many tide gauges records in Japan do not start before 1969). 
We then investigate distributional shifts for tide gauges with more than 80 years of data and less than 20$\%$ of missing values, which amounts to 28 available tide gauges. Such long periods translate in a more robust statistical inference of changes in quantiles. For all tide gauges, the daily climatology is removed from each record by subtracting to each day its average over the whole period of interest. Details on the tide gauges preprocessing are further discussed in the Appendix, Section \ref{app:preprocessing}. Start and end recording dates of tide gauges with records longer than 80 years are presented in the Appendix, Section \ref{app:long_records_tide_gauges}.

\subsection{GFDL-CM4 Climate model: strengths and limitations}

We consider outputs from the GFDL-CM4 model \citep{Held}. The ocean component is the MOM6 ocean model \citep{OM4} with horizontal grid spacing of 0.25$^\circ$ and 75 vertical layers. With this grid spacing, the model cannot resolve many coastal bays and harbors. Nonetheless, the horizontal grid spacing of 0.25$^\circ$ allows to realistically represent the sloping of continental shelves and its sharp deepening at the boundaries with the open oceans. Such important feature cannot be represented in traditional 1$^\circ$ coupled models. Accurate representation of coastal geometry is key to trustworthy simulations of storm surges and coastal sea level as discussed in \cite{jianjun}.\\

The atmospheric/land component is the AM4 model \citep{AM4a,AM4b} with a horizontal grid spacing of roughly 1$^\circ$ and 33 vertical layers. Despite its relatively coarse grid spacing, the model simulates the physical characteristics of tropical cyclones reasonably well, allowing one to study their frequency and changes under different forcings \citep{AM4a}. However, strong hurricanes (i.e., category 4 and 5) are not simulated, hence their impact on sea level extremes cannot be assessed in our study \citep{jianjun,AM4a}.\\

The contribution to sea level rise caused by melting of land ice is missing in CM4 due to the lack of an ice-sheet model. Note that this is the case for all coupled models in the Coupled Model Intercomparison Project-phase 6 (CMIP6) \citep{cmip6}. Tides are not simulated and the associated tide surges \citep{Rego}, driven by constructive interactions of tides and storm surges, are absent from CM4. Climate-unrelated factors such as glacial isostatic adjustment and terrestrial water storage, both relevant to sea level (mean) trends \citep{Wang}, are also not simulated by this model. As shown by \cite{Winton}, CM4 is a high climate sensitivity model, and so it is likely too sensitive to anthropogenic forcing. A thorough investigation of CM4-model biases in terms of sea level variability was presented in \cite{jianjun}. Despite some of its limitations, the authors showed that the GFDL-CM4 model can serve as a useful framework to explore changes in sea level statistics under different forcing scenarios. It thus directly serves our interests in the present study.\\ 

The focus here is on two simulations. First, we consider one historical experiment in CM4, from 1970 to 2014 with external forcings consistent with observations. Second, we focus on an idealized experiment with a 1$\%$ $\text{CO}_{2}$ increase per year (hereafter ``1pctCO2''). This experiment simulates the climate system under a 1$\%$ increase of $\text{CO}_{2}$ per year for 150 years; from preindustrial global $\text{CO}_{2}$ concentrations up to quadrupling at year 140. We assume that the anthropogenic signal of sea level rise emerges during the first 50 years of the simulation (as shown by \cite{jianjun} for the US east coast) and focus only on the last 100 years. The 1pctCO2 experiment is especially useful since under an incremental, yearly increase in $\text{CO}_{2}$ concentration, sea level statistics are expected to change almost linearly (at least in the last 100 years). This linear change allows us to leverage the framework proposed in Section \ref{sec:methodology}, and to explore the emergence of changes in distributional shapes in coastal sea levels.\\

The model output is remapped to a uniform 0.5$^\circ$ grid and only grid cells in the latitudinal range $[-60^\circ,60^\circ]$ are considered. In both simulations, outputs are daily averages. We focus only on coastal locations, accounting for 6318 time series in each case.

\subsection{Sea level decomposition}

Following \cite{Gill}, \cite{Yin_etal_AR4_2010}, and \cite{GRIFFIES201237}, we decompose the sea level time tendency in a hydrostatic and Boussinesq ocean  according to the following sea level budget 
\begin{linenomath}
\begin{equation}
    \Delta \eta = 
      \frac{\Delta P_{b}}{\rho_{0} \, g}
     -\frac{\Delta {P}_{a}}{\rho_{0} \, g}
     - \frac{1}{\rho_{0}} \int_{-H}^{\eta} \Delta \rho \, \mathrm{d}z,
\label{eq:sea_level_decomposition}
\end{equation}
\end{linenomath}
where $\Delta$ refers to a temporal increment relative to an initial time. Variables $\eta, P_{b}, P_{a}$ and $\rho$ respectively represent the sea level, bottom and surface pressure and density at time $t$ and longitude, latitude, vertical position $(x,y,z)$. We make the Boussinesq approximation, which accounts for the constant reference density $\rho_{0}$ (e.g., Section 2.4 of \cite{Vallis2017}).\\  

In equation \eqref{eq:sea_level_decomposition}, $P_{b}$ is the ocean bottom pressure, so that $\Delta P_{b}/(\rho_{0} \, g)$ measures the change in sea level associated with changes in water column mass. Changes in water column mass arise from the convergence of mass via ocean currents as well as water crossing the ocean free surface via precipitation, evaporation, river runoff, and sea ice formation/melt. The sea level pressure, $P_{a}$, is caused by atmospheric and sea ice loading. The contribution from $P_{a}$ is referred to as the inverse barometer (e.g., \cite{Ponte}), whereby increases in $P_{a}$ lead to a decrease in sea level. The final term in the budget \eqref{eq:sea_level_decomposition} is the local steric effect, which is measured by the depth integral of changes to {\it in situ} ocean density. For example, a decrease in column integrated density, such as through ocean warming or freshening, leads to a sea level rise from local steric expansion.\\

Changes from bottom pressure and the local steric effect are associated with ocean dynamics and ocean density, and are referred to as sterodynamic sea level changes \citep{gregory}. Following the Coupled Model Intercomparison Project Phase 6 (CMIP6) convention detailed in Appendix H of \cite{OMIPa_2016}, the sea level diagnostic ``zos'' measures the regional sterodynamic changes by setting the global mean to zero at each diagnostic time step, with \cite{OMIPa_2016} referring to \textit{zos} as the dynamic sea level. There have been many studies of changes to the global mean sea level (e.g., \citep{frederikse2020causes,Palmer,Bittermann,Bars} among many others), with \cite{jianjun} discussing the global thermosteric sea level rise in the CM4 model used here. When  comparing models to observations, the global thermosteric sea level must be added to the dynamic sea level at every location in the ocean. However, our interest concerns changes in higher order moments that are independent of changes in the global mean sea level. Therefore, we focus our analysis on the dynamic sea level and inverse barometer only.\\

We furthermore focus on the anomaly patterns of the inverse barometer (i.e, we remove the daily climatology). Note that the inverse barometer contribution is not explicitly simulated by CMIP5 models. We thus  diagnose this term offline by saving the sea level pressure from the atmospheric model. As for tide gauges, we remove the climatological daily cycle from all fields before performing our analysis.

\section{Methodology} \label{sec:methodology}

In this work we aim to quantify linear changes in sea level distributions in the observational records and climate model output. The methodology is general and can be potentially applied to any climate observable with close-to-linear changes in distributions.\\

\begin{figure}[tbhp]
\centering
\includegraphics[width=0.8\linewidth]{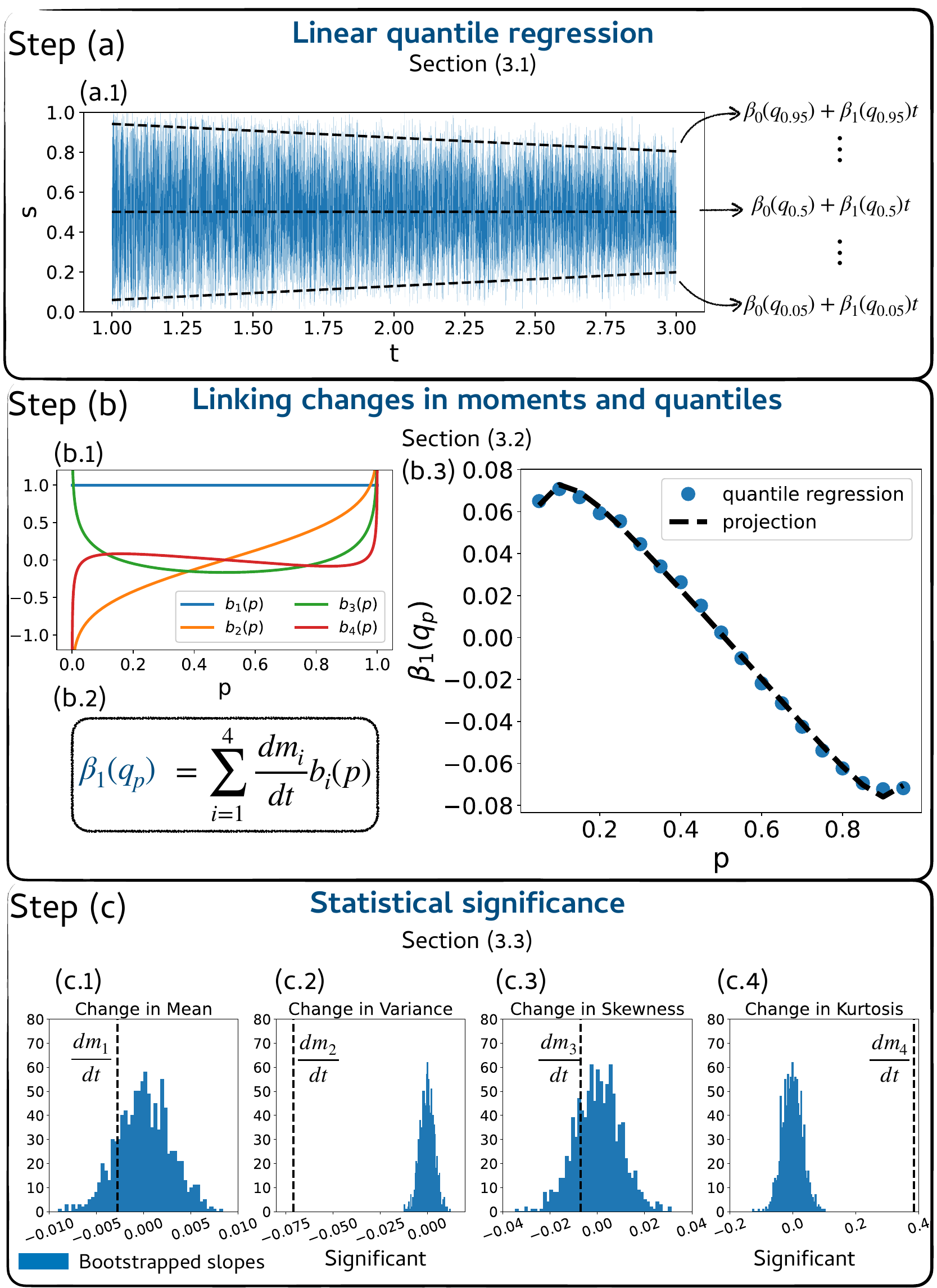}
\caption{Schematic of the proposed framework illustrated using a synthetic time series. We generate a time series $\{(t_{1},s_{1}),(t_{2},s_{2}),...\}$ sampled from a time dependent Beta distribution $P(s,t)$ (see Section \ref{subsec:synthetic}). Temporal changes in statistical moments are computed analytically {\it a priori} and are chosen to come solely from the variance (second moment) and kurtosis (fourth moment). Step (a): we apply quantile regression for the range of quantiles $q_{p}$, with $p \in [0.05,0.95]$ every $dp = 0.05$ for a total of $19$ slopes. In panel (a) we show the quantile regression for $p = 0.95, 0.5$ and $0.05$. Note that $q = q(p)$ and $q_{p}$ is used only for convenience. Step (b): we project the 19 quantile slopes onto a set of four orthogonal polynomials (see panel b.3). Step (c): we quantify the statistical significance of coefficients $\frac{d m_{i}}{dt}$ quantifying how \textit{independent} changes in moments explain changes in quantiles computed in step (a) ($\frac{d m_{i}}{dt}$; with $i\in[1,4]$ representing changes in mean, variance, skewness and kurtosis, respectively). Note that in Section \ref{sec:significance_test} we refer to $\frac{d m_{i}}{dt}$ as $a_{i}$ to simplify the notation. Significant changes at the $95\%$ level come for this synthetic time series solely from the second and fourth moments, as known from analytical results. Additional synthetic tests are presented in the Section \ref{subsec:synthetic}}
\label{fig:schematic}
\end{figure}

Given a sea level time series, we apply quantile regression to measure linear trends in quantiles \citep{Koenker,Portnoy,koenker2001quantile}.
We then introduce a set of orthogonal functions and project the quantile slopes onto such functions, in order to quantify how changes in quantiles are driven by changes in the first four statistical moments. Crucially, slopes in the proposed polynomials are orthogonal to each other by construction, allowing us to decouple different sources of distributional changes. Here, we present the three main steps in the framework: (a) quantile regression, (b) projection onto polynomials, (c) statistical significance test. A schematic of the proposed framework is presented in Figure \ref{fig:schematic}.

\subsection{Quantile Regression}
\label{subsec:quantile-regression}

A quantile function $Q_{X}(p), p \in [0,1]$, of a random variable $X$ returns a value $x$ such that $(p \times 100) \%$ of the values are less than $x$. For example, the 0.95 quantile $Q_{X}(0.95)$ is a real number $x$ such that $95\%$ of the values of the random variable, $X$, are smaller than $x$. $Q_{X}(p), p \in [0,1]$ is the inverse $F^{-1}_{X}(p)$ of the cumulative distribution function $F_{X}(x) = P(X\leq x) = p$, representing the probability $p$ that $X$ would take a value less than or equal to $x$.  In what follows we simplify the notation and refer to $Q_{X}(p)$ as $q_{p}$ and to cumulative functions as $F(x)$.\\

Quantile regression allows  us to estimate temporal changes in the conditional \textit{median} (i.e., $q_{0.5}$) or any other quantile of a time dependent  distribution \citep{Koenker,koenker2001quantile}. Here, we focus on the case of linear regression. For each quantile, $q_{p}$, we need to fit two parameters: an intercept $\beta_{0}(q_{p})$ and a slope $\beta_{1}(q_{p})$ (see Figure \ref{fig:schematic}). Differently from linear regression for which we minimize the mean square error, the quantile regression process places asymmetric weights on positive and negative residuals \citep{koenker2001quantile}.\\

Formally, given $n$ observations at times $t_{i}$, written as  $\{(t_{1},s_{1}),(t_{2},s_{2}),...\}$, and the assumption of linear relationship for each quantile  $q_{p}(t), p \in [0,1]$, then the goal is to minimize the following cost function:
\begin{linenomath}
\begin{equation}
\underset{\beta_{0}(q_{p}),\beta_{1}(q_{p}) \in \mathbb{R}}{\arg \min} \sum_{i=1}^{n}\rho_{p}(s_{i}-\beta_{0}(q_{p})-\beta_{1}(q_{p}) \, t_{i}),
\label{eq:quantile_regression}
\end{equation}
\end{linenomath}
where the ``check function'' $\rho_{p}(u) = p \max(u,0) + (1-p) \max(-u,0)$, with $p \in [0,1]$,  assigns asymmetric weights to residuals.
For a given time series, we apply quantile regression for the range of quantiles $q_{p}$ with $p \in [0.05,0.95]$ every $dp = 0.05$ for a total of $19$ slopes (as done in \cite{mckinnon2016changing}). This number of slope was found to be sufficient to discover the right changes in moments, as shown later on in the next section.  An example of quantile regression for $p = 0.95, 0.5$ and $0.05$ is shown in Figure \ref{fig:schematic}(a.1).\\

In practice, quantile regression studies depend on the existence of very fast algorithms \citep{Chernozhukov}. \cite{Portnoy} proposed a precise and time-efficient minimization method based on the Frisch-Newton algorithm. However, the computation time is still rather long in the case of multiple quantile regressions and bootstrap inference. To this end, very recently \cite{Chernozhukov} showed that the computation of many quantile regressions can be (vastly) accelerated by exploiting nearby quantile solutions. This method was  essential for the completion of our work, especially when analyzing the climate model outputs, which included  12636 time series each with 36500 time steps. The algorithm can be found in the quantile regression package ``quantreg''  implemented in \textsf{R} as ``quantile regression fitting via interior point methods'' (i.e., method \textsf{pfnb}) \citep{quantreg}. In the case of sea level each quantile $q_{p}$ has dimensionality of $[q_{p}] = [\text{length}]$, so that the slopes have dimensions of $[\beta_{1}(q_{p})] = [\frac{\text{length}}{\text{time}}]$.

\subsection{Linking changes in quantiles to changes in statistical moments}
\label{subsec:basis-functions}

The quantile regression step allows us to quantify slopes in $N$ quantiles. This quantification can come at the expense of interpretability, especially when $N$ is large and when analyzing more than one time series. To facilitate analysis and interpretation, we quantify how changes in quantiles of a distribution are driven by changes in the mean, variance, skewness and kurtosis therefore defining a single framework to study changes in quantiles and moments. All throughout the paper we refer to the mean, variance, skewness and kurtosis as $(m_{1},m_{2},m_{3},m_{4})$ for practical convenience, and report their mathematical expression in the Appendix, Section \ref{app:moments_definition}. A possible way to link changes in quantile  to changes in moments was proposed in \cite{mckinnon2016changing}. The authors empirically derived a set of polynomials by observing how quantiles of a distribution change when changing its moments one at a time. The lack of orthogonality of such functions motivated the authors to adopt Legendre polynomials, which are orthogonal by construction and share similar shapes to the derived functions. Inspired by \cite{mckinnon2016changing}, here we approach the problem from a theoretical point of view. We take advantage of the work of \cite{Cornish} and define a unified framework useful to study and explore temporal changes in both quantiles and moments of a distribution.

\subsubsection{Stationary process}
\label{subsubsec:Cornish-Fisher}
The starting point of our derivation is the Cornish-Fisher Expansion \citep{Cornish,Fisher,Wallace,Hill}. \cite{Cornish} derived an asymptotic expansion expressing any quantile of a distribution as a function of its cumulants.\footnote{Cumulants are quantities providing an alternative to moments of a distribution. A description based on cumulants often simplifies theoretical studies in statistics and probability theory. In this paper we focus on moments of the distribution and refer the interested reader to \cite{Cornish,Kendall} for more details on cumulants.} Studies often focus on a modified version written in terms of the first four distributional moments. Furthermore, in practical applications higher order moments show large sensitivity to sampling fluctuations and \cite{Bekki} showed that a fourth order truncation often allows for very accurate estimations of quantiles.\\

Given a stationary distribution with mean $m_{1}$, variance $m_{2}$, skewness $m_{3}$, and excess kurtosis $m_{4}$ parameters, the following asymptotic relation holds \citep{Cornish,Bekki}:
\begin{linenomath}
\begin{equation}
q_{p} \sim m_{1} + \sqrt{m_{2}} \, w;~ w = z_{p} + (z_{p}^{2}-1)\frac{m_{3}}{6} + (z_{p}^{3}-3z_{p})\frac{m_{4}}{24}-(2z_{p}^{3}-5z_{p})\frac{m_{3}^{2}}{36},
\label{eq:cornish_fisher}
\end{equation}
\end{linenomath}
where $q_{p}$ is the quantile of the ``true'' distribution and $z_{p}$ is the quantile of a standard normal $\mathcal{N}(0,1)$ at $p\in[0,1]$\footnote{We remind the reader that $q_{p}=Q_{X}(p)$ is the quantile function of a random variable $X$. We write $q_{p}$ to simplify the notation. The function $z_{p}$ is the quantile function of a standard normal $\mathcal{N}(0,1)$. In Python $z_{p}$ is given by  \tiny{\textsf{scipy.stats.norm.ppf(p, loc=0, scale=1)}}.}. 
In other words, $z_{p}$ is the inverse $F^{-1}(p)$ of the cumulative distribution $F(x)=(2\pi)^{-1/2}\int_{-\infty}^{x}\exp{(-\upsilon^{2}/2)}d\upsilon$ of a Gaussian distribution with zero mean and unit variance. Importantly, $q_{p}$, $m_{1}$ and $\sqrt{m_{2}}$ have the same dimension, whereas $z_{p}$, $m_{3}$ and $m_{4}$ are non-dimensional. A synthetic and a first climate-related test are discussed in the Appendix, Section \ref{app:CF_testing}.\\

In the case of a normal distributions, $m_{3} = m_{4} = 0$ and the expansion \eqref{eq:cornish_fisher} reduces to $q_{p} = m_{1} + \sqrt{m_{2}} \, z_{p}$. The inclusion of the third and fourth moments, $m_{3}$ and $m_{4}$, allows us to approximate any quantile of a non-normal distribution as a function of a normal (Gaussian) distribution. This formula is a powerful tool as it can provide asymptotic estimations of arbitrarily large quantiles, otherwise difficult from data alone. It is useful for distributions that do not show large differences from a normal distribution with the domain of validity further discussed by \cite{Maillard} and \cite{Manesme}.

\subsubsection{Non-stationary process}
\label{subsubsec:derivation}
Here, we aim to understand and quantify how temporal changes in individual quantiles are driven by changes in statistical moments. In theory, each quantile $q_{p}$ is time dependent through all moments of the distribution. Here we consider dependence only up to the first four moments:
\begin{linenomath}
\begin{equation}
q_{p}(t) = q_{p}(m_{1}(t),m_{2}(t),m_{3}(t),m_{4}(t)).
\label{eq:quantile_time_dependence}
\end{equation}
\end{linenomath}
Here $m_{1}, m_{2}, m_{3}$ and $m_{4}$ refer to the mean, variance, skewness and excess kurtosis. We focus on linear changes in time and therefore write
\begin{linenomath}
\begin{equation}
\frac{dq_{p}}{dt} = \frac{\partial q_{p}}{\partial m_{1}} \frac{dm_{1}}{dt} + \frac{\partial q_{p}}{\partial m_{2}} \frac{dm_{2}}{dt} + \frac{\partial q_{p}}{\partial m_{3}} \frac{dm_{3}}{dt} + \frac{\partial q_{p}}{\partial m_{4}} \frac{dm_{4}}{dt} = \sum_{i=1}^{4}\frac{\partial q_{p}}{\partial m_{i}}\frac{d m_{i}}{dt}.
\label{eq:quantile_in_time}
\end{equation}
\end{linenomath}
Each term $\frac{\partial q_{p}}{\partial m_{i}}$ in Eq. \eqref{eq:quantile_in_time} can be further evaluated by differentiating Eq. \eqref{eq:cornish_fisher}:
\begin{linenomath}
\begin{align}
\begin{cases}
\frac{\partial q_{p}}{\partial m_{1}} &= 1
\\
\frac{\partial q_{p}}{\partial m_{2}} &= \frac{1}{2}\frac{1}{\sqrt{m_{2}}}[z_{p} + \frac{1}{6}(z_{p}^{2}-1) \, m_{3} - \frac{1}{36}(2 z_{p}^{3} -5z_{p}) \, m_{3}^{2}+\frac{1}{24}(z_{p}^{3} -3z_{p}) \, m_{4}]
\\
\frac{\partial q_{p}}{\partial m_{3}} &= \sqrt{m_{2}} \, [\frac{1}{6}(z_{p}^{2}-1) - \frac{1}{18}(2 z_{p}^{3} -5z_{p}) \, m_{3}]
\\
\frac{\partial q_{p}}{\partial m_{4}} &= \frac{\sqrt{m_{2}}}{24} (z_{p}^{3} -3z_{p}).
\end{cases}
\label{eq:derivatives}
\end{align}
\end{linenomath}
We assume relatively small deviation from Gaussian behaviour and focus only on first-order changes in Eq.~\eqref{eq:derivatives}. This is a reasonable assumption as the dynamics of many climate observables often show a strong Gaussian component, especially when focusing on anomalies (after removing climatologies). In case of (relatively) large deviation from Gaussianity, the framework is still useful at first order as suggested by tests in Appendix \ref{app:beta_test}. In other words we evaluate Eq.~\eqref{eq:derivatives} \textit{locally} at the point $m_{*} = (m_{1} = 0, m_{2} = 1, m_{3} = 0, m_{4} = 0)$ to describe changes in the neighborhood of a Gaussian distribution. Therefore, we relate the slopes $\beta_{1}(q_{p})$ computed in the quantile regression step in Eq. \eqref{eq:quantile_regression} to changes in moments as: 
\begin{linenomath}
\begin{equation}
\beta_{1}(q_{p}) \sim \frac{dq_{p}}{dt}\Big|_{m_{*}} = \sum_{i=1}^{4}\frac{d m_{i}}{dt}\frac{\partial q_{p}}{\partial m_{i}}\Big|_{m_{*}} = \sum_{i=1}^{4}\frac{d m_{i}}{dt} b_{i}(p).
\label{eq:projection}
\end{equation}
\end{linenomath}
Here the set of polynomials, $b_{i}(p)$, are defined by evaluating the terms $\frac{\partial q_{p}}{\partial m_{i}}$; $i\in[1,4]$ in Eq. \eqref{eq:derivatives} in the point $m_{*} = (m_{1} = 0, m_{2} = 1, m_{3} = 0, m_{4} = 0)$:
\begin{linenomath}
\begin{align}
\begin{cases}
b_{1}(p) = \frac{\partial q_{p}}{\partial m_{1}}|_{m_{*}} &= 1\\
b_{2}(p) = \frac{\partial q_{p}}{\partial m_{2}}|_{m_{*}} &= \frac{z_{p}}{2}\\
b_{3}(p) = \frac{\partial q_{p}}{\partial m_{3}}|_{m_{*}} &= \frac{1}{6}(z_{p}^{2}-1)\\
b_{4}(p) = \frac{\partial q_{p}}{\partial m_{4}}|_{m_{*}} &= \frac{1}{24} (z_{p}^{3} -3z_{p}).\\
\end{cases}
\label{eq:derivatives_in_the_neighborhood}
\end{align}
\end{linenomath}
Such functions are scaled Hermite polynomials of the function $z_{p}$ and defined in the range $p \in [0,1]$. The  polynomials in  Eq.~\eqref{eq:derivatives_in_the_neighborhood} 
satisfy 
\begin{linenomath}
\begin{align}
\int_{0}^{1} b_{i}(p) b_{j}(p) dp = 0 \quad  \text{if}~ i \neq j,
\label{eq:orthogonality}
\end{align}
\end{linenomath}
and so they are orthogonal to each other. We depict these four functions in Figure \ref{fig:schematic}(b.1).\footnote{We also tested the impact of further rescaling the polynomials to define an orthonormal rather than just orthogonal set. In the orthonormal case, all synthetic tests give the same results in terms of statistical significance but do not capture the relative scaling between changes in moments.}\\

Eq.~\eqref{eq:projection} allows us to examine how time changes in distributional quantiles are driven by time changes in the first four moments. We note that apart from $b_{1}(p)$, all polynomials in Eq. \eqref{eq:derivatives_in_the_neighborhood} differ from the ones in \cite{mckinnon2016changing}. As an example, an important difference is in the $b_{2}(p)$ function, quantifying changes in variance. In the case of a drifting gaussian distribution, $b_{2}(p)$ quantifies the \textit{exact} link between changes in moments and quantiles. In the ``bulk'' of the distribution, the function $b_{2}(p)$ derived in Eq.~\eqref{eq:derivatives_in_the_neighborhood} gives negative(positive) corrections to the correspondent function in \cite{mckinnon2016changing} for $p$ greater(smaller) than $p = 0.5$. Furthermore, $b_{2}(p)$ shows how changes in variance also drive changes in distributional tails, with quantiles diverging to $+\infty(-\infty)$ when $p\rightarrow(1)0$.\\

The slopes $\beta_{1}(q_{p})$ appearing in Eq. \eqref{eq:projection} can be efficiently estimated through the quantile regression step as shown in Eq. \eqref{eq:quantile_regression}. The polynomials $b_{i}(p)$ have been derived in Eq. \eqref{eq:derivatives_in_the_neighborhood} and follows from \cite{Cornish}. Therefore, changes in moments, $\frac{d m_{i}}{dt}$, can be computed as the least-squares solution of Eq. \eqref{eq:projection}. An example is shown in Figure \ref{fig:schematic}(b.3) for a Beta distribution with changes coming exclusively from variance and kurtosis (i.e., $m_{2}$ and $m_{4}$). $\frac{d m_{1}}{dt}$ is equivalent to a linear regression. We refer to the coefficients $\frac{d m_{i}}{dt}$ as changes (or slopes) in moments. However, such inferred changes are orthogonal to each other, and this property may not be the case for changes in the \textit{true} moments. The orthogonality constraint offers a useful and interpretable way to decompose \textit{independent} sources of shifts in a distribution.\\

In the case of sea level, the dimensionality in changes in moments $\frac{d m_{i}}{dt}$; $i \in [1,4]$ is as follows: $[\frac{d m_{1}}{dt}] = [\frac{\text{length}}{\text{time}}], [\frac{d m_{2}}{dt}] = [\frac{\text{length}^2}{\text{time}}], [\frac{d m_{3}}{dt}] = [\frac{1}{\text{time}}], \frac{d m_{4}}{dt} =  [\frac{1}{\text{time}}]$.

\subsection{Statistical significance} \label{sec:significance_test}

The significance test is performed via the bootstrapping methodology \citep{Chernick}.
For a given dataset (i.e., model's output or tide gauges), we consider the $j$-th time series, $x_{j}(t)$, and estimate the statistical significance of each coefficient $a_{i} = \frac{d m_{i}}{dt}$, $i \in [1,4]$. We do so by permuting (with replacement) $x_{j}(t)$ $B$ times and recomputing the slopes in moments $a_{i}$ at each iteration. We choose a value of $B = 1000$ and show the sensitivity of such a choice for one time series in Appendix \ref{app:balboa_record}, Figure \ref{fig:bootstrap_tests} (results do not change on average for randomly chosen time series). This method allows to construct a distribution approximating the null-distribution under the null hypothesis of stationarity. For this re-sampling approach we use block-bootstrapping, with blocks of one season to satisfy the assumption of statistical independence required in the bootstrapping test \citep{Chernick}. The robustness on this choice of block size is mentioned in Section \ref{sec:results} and detailed in the Appendix, Section \ref{app:long_records_tide_gauges}. The specific block-bootstrapping considered is the moving block-bootstrapping that allows for overlapping blocks as described in \cite{kunsch}.\\

We consider the following two significance tests. 

\begin{enumerate}

    \item When dealing with one or few time series, we consider a two-tailed test and deem slopes as significant at level $\alpha$ if in the $(1- \frac{\alpha}{2}) \%$ (or $\frac{\alpha}{2} \%$) of the upper (or lower) tails of the bootstrapped distribution. We choose $\alpha = 0.05$ (i.e., $95\%$ significance level).
    
    \item When considering many time series we face a multiple testing problem and more and more false positives are expected in the statistical inference \citep{Witten}. To control the ratio of false positives we adopt the False Discovery Rate (FDR) test proposed by \cite{Benjamini}. For a given time series $x_{j}(t)$ and a corresponding slope $a_{i}$, $i \in [1,4]$, we compute a p-value $p_{i}$. We perform this computation for all $M$ time series in the dataset under investigation. We then rank all $M$ p-values in ascending order and keep only the first $m < M$ so that $p_{m}$ is the largest p-value such that $p_{m} < \phi\frac{m}{M}$, where $\phi$ is the False Discovery Rate (FDR). Note that the False Discovery Rate is usually denoted by the letter ``$q$''. Here we use $\phi$ to avoid confusions with quantiles $q$ in the quantile regression step. In plain words, a FDR of $10\%$ would imply that no more than $10\%$ of the rejected null hypotheses are False Positives. This method has been shown to be an efficient test to control \textit{on average} the ratio of false positives arising from multiple testing \citep{Benjamini} and found applications in climate science \citep{Ventura,Wilks,Ilias1,Runge}.\\
    
    In our case, this step requires to approximate a p-value from a bootstrapped distribution. Given the $j$-th time series $x_{j}(t)$ and a slope $a_{i} = \frac{dm_{i}}{dt}$, $i \in [1,4]$, we denote $a_{i}^{*1}, a_{i}^{*2}, ..., a_{i}^{*B}$ as the $B$ bootstrapped slopes. We define the correspondent p-value of $a_{i}$ as $p_{i} = \frac{\sum_{b=1}^{B}1_{\mid a_{i}^{*b} \mid \geq \mid a_{i} \mid}}{B}$; where $1_{\mid a_{i}^{*b} \mid \geq \mid a_{i} \mid} = 1$ if $\mid a_{i}^{*b} \mid\geq \mid a_{i} \mid$ and zero otherwise \citep{Witten}.
    
\end{enumerate}

\subsection{Synthetic tests} \label{subsec:synthetic}

\begin{figure}[tbhp]
\centering
\includegraphics[width=1\linewidth]{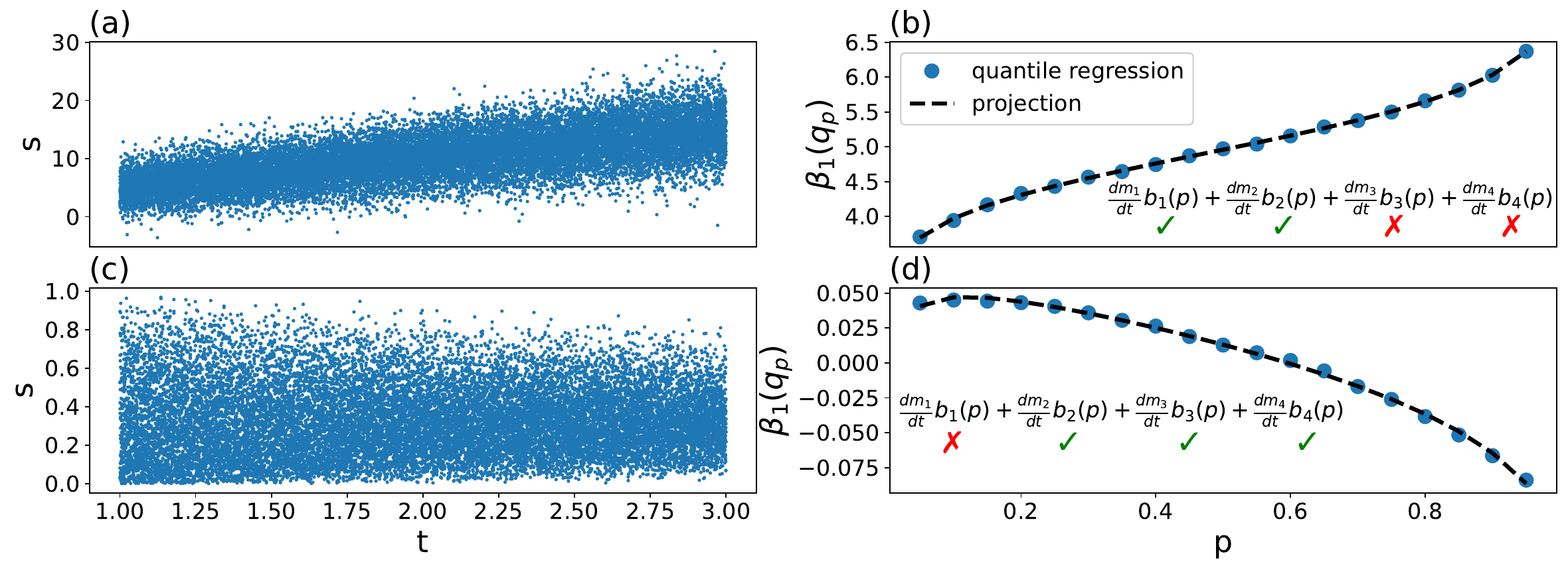}
\caption{Testing the methodology on time series $\{(t_{1},s_{1}),(t_{2},s_{2}),...\}$ sampled from time dependent Gaussian and Beta distributions $P(s,t)$. Panel (a): time series sampled from a time-dependent Gaussian distribution. Panel (b): linear quantile slopes for the time series shown in panel (a). The slopes $\beta_{1}(q_{p})$ are computed for $p \in [0.05, 0.95]$ every $dp = 0.05$ and indicated as blue dots. The black, dashed line indicates the projection onto polynomials as defined in Eq. \eqref{eq:projection}. Green ``check'' marks indicate statistically significant ($95\%$ confidence level) changes in moments; red "crosses" indicate non-significant changes. Panel (c): time series sampled from a time-dependent Beta distribution. Panel (d): same as panel (b) but for the case of the Beta distribution. In both cases, the method identifies the statistical moments driving changes in the distributions}
\label{fig:testing}
\end{figure}

In this section we test the proposed framework on two time-depedent distributions with known changes in moments. We consider two time-dependent processes sampled from a Gaussian and Beta distribution and display tests in Figure \ref{fig:testing}.

\begin{itemize}
    \item We define a time-dependent Gaussian process $P(x,t) = \frac{1}{\sqrt{2 \pi b t}}e^{\frac{(x-a t)^{2}}{2 b t}}$ defined over $x \in \mathbb{R}$ with mean $m_{1} = a t$ and variance $m_2 = b t$. $a, b \in \mathbb{R}$ and $t$ represents the time vector. We choose $a=b=5$. In this case we expect distributional changes to be driven exclusively by the mean and variance.
    \item We consider a stationary Beta distribution $P(x)=\frac{x^{\alpha -1}(1-x)^{\beta-1}}{B(\alpha,\beta)}$ defined over $x \in [0,1]$. Here $B(\alpha,\beta) = \frac{\Gamma(\alpha)\Gamma(\beta)}{\Gamma(\alpha+\beta)}$, with $\Gamma$ being the Gamma function \citep{PhilipDavis}. Therefore, $P(x)$ depends on two parameters, $\alpha$ and $\beta$, controlling the ``thickness'' of the tails of the distribution. A larger $\alpha$ ($\beta$) implies negative (positive) skewness; the distribution is symmetric if $\alpha = \beta$. The mean, variance, skewness and kurtosis read as:
    \begin{linenomath}
    \begin{align}
    \begin{cases}
    m_{1} &= \frac{\alpha}{\alpha + \beta}\\
    m_{2} &= \frac{\alpha \beta}{(\alpha + \beta)^{2}(1+\alpha + \beta)}\\
    m_{3} &= \frac{2(\beta-\alpha)\sqrt{1+\alpha+\beta}}{\sqrt{\alpha}\sqrt{\beta}(2+\alpha+\beta)} \\
    m_{4} &= \frac{3(1+\alpha+\beta)(\alpha\beta(\alpha+\beta-6)+2(\alpha+\beta)^{2})}{\alpha\beta (2+\alpha+\beta) (3+\alpha+\beta)}.
    \label{eq:moments_beta}
    \end{cases}
    \end{align}
    \end{linenomath}
    We then introduce a ``drift'' in this distribution by prescribing $\alpha = a t$ and $\beta = b t$, and choose $a = 1 $ and $ b = 2$. This defines a process with constant mean, but time dependent variance, skewness and kurtosis. For these parameters the exact moments are then $(m_{1},m_{2},m_{3},m_{4})=(\frac{1}{3},\frac{2}{9+ 27 t},\frac{\sqrt{2 + 6t}}{2+3t},3 - \frac{3}{2 + 3 t})$.
\end{itemize}

In both cases we create a time-dependent process $\{(t_{1},s_{1}),(t_{2},s_{2}),...\}$ in the temporal range $t \in [1,3]$. In this temporal range, changes can be weakly nonlinear but still far from strong nonlinearities. Specifically, at each time step $dt = 10^{-4}$ we random sample from the two distributions $P(s,t)$. We then apply the proposed framework to (a) estimate quantile slopes $\beta_{1}(q_{p})$ and (b) compute the respective changes in moments $\frac{d m_{i}}{dt}$. By random sampling at each $dt$, consecutive time steps are independent by definition and block sizes in the boostrapping procedure are equal to 1 time step. Results for the Gaussian and Beta distributions cases are shown in Figure \ref{fig:testing}. The method correctly identifies the statistical moments driving changes in the distributions. We note that the proposed \textit{linear} framework is still valid in the presence of weak nonlinearities such as for $m_{2},m_{3}$ and $m_{4}$ for the process sampled from the Beta distribution.\\

Tests for a process with prescribed changes in only the variance, $m_{2}$, and kurtosis, $m_{4}$, are shown in the schematic in Figure \ref{fig:schematic}.

\clearpage

\section{Results} \label{sec:results}

\subsection{Changes in coastal sea level distributions from tide gauges} \label{sec:all_tide_gauges} 

\begin{figure}[tbhp]
\centering
\includegraphics[width=1\linewidth]{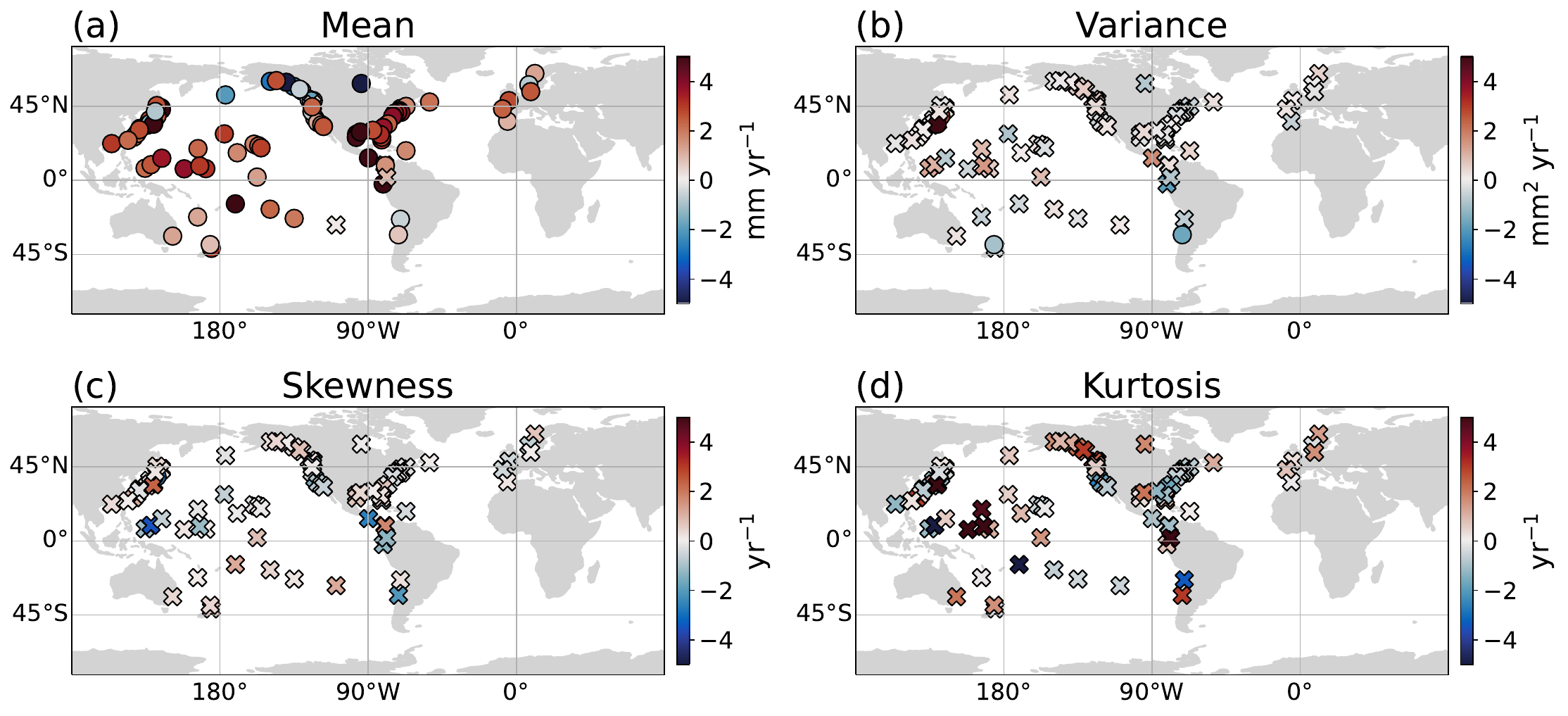}
\caption{Projection onto mean, variance, skewness and kurtosis polynomials for the period 1970-2017. Statistically significant changes are marked with filled ``circles'', whereas  insignificant changes are marked with ``X''s. Statistical significance is computed using the FDR test with threshold $\phi = 0.05$}
\label{fig:fdr_tide_gauges_1970_2017}
\end{figure}

Changes in sea level distributions measured by tide gauges for the period 1970-2017 are summarized in Figure \ref{fig:fdr_tide_gauges_1970_2017} and in Table \ref{table:significance_obs}. The statistical significance test used is the False Discovery Rate (FDR, see Section \ref{sec:significance_test}) with $\phi = 0.05$. The p-values are computed from the block-bootstrapped distributions (Section \ref{sec:significance_test}) and a sensitivity analysis on the chosen block-size is shown in the Appendix, Section \ref{app:p-values-tide-gaues}.\\

The main significant changes in the observed coastal sea level distributions come from a shift in the mean. Contributions coming from higher-order moments are less significant and sporadic (see Table \ref{table:significance_obs}). Among the 92 tide gauges with significant changes in the mean, the average slope is $2.08$~mm/yr (see Table \ref{table:significance_obs}). However, if only the positive significant slopes are considered ($78$ of the $92$ tide gauges) then the slope increases to $2.92$~mm/yr.\\

Drivers of mean changes in tide gauges have been discussed in \cite{Wang}, where the authors identified  stereodynamic sea level changes (i.e., changes driven by currents, temperature and salinity) as the main contributors of the observed trends. Large trends in the US east coast mainly come from the combined effect of sterodynamic and Glacial Isostatic Adjustments (GIA) contributions. Downward sea level trends are driven mainly by GIA (e.g., Baltic Sea \citep{Weisse}), terrestrial water storage or mass loss of land ice (e.g. Alaska coastline).\\

\begin{figure}[tbhp]
\centering
\includegraphics[width=1\linewidth]{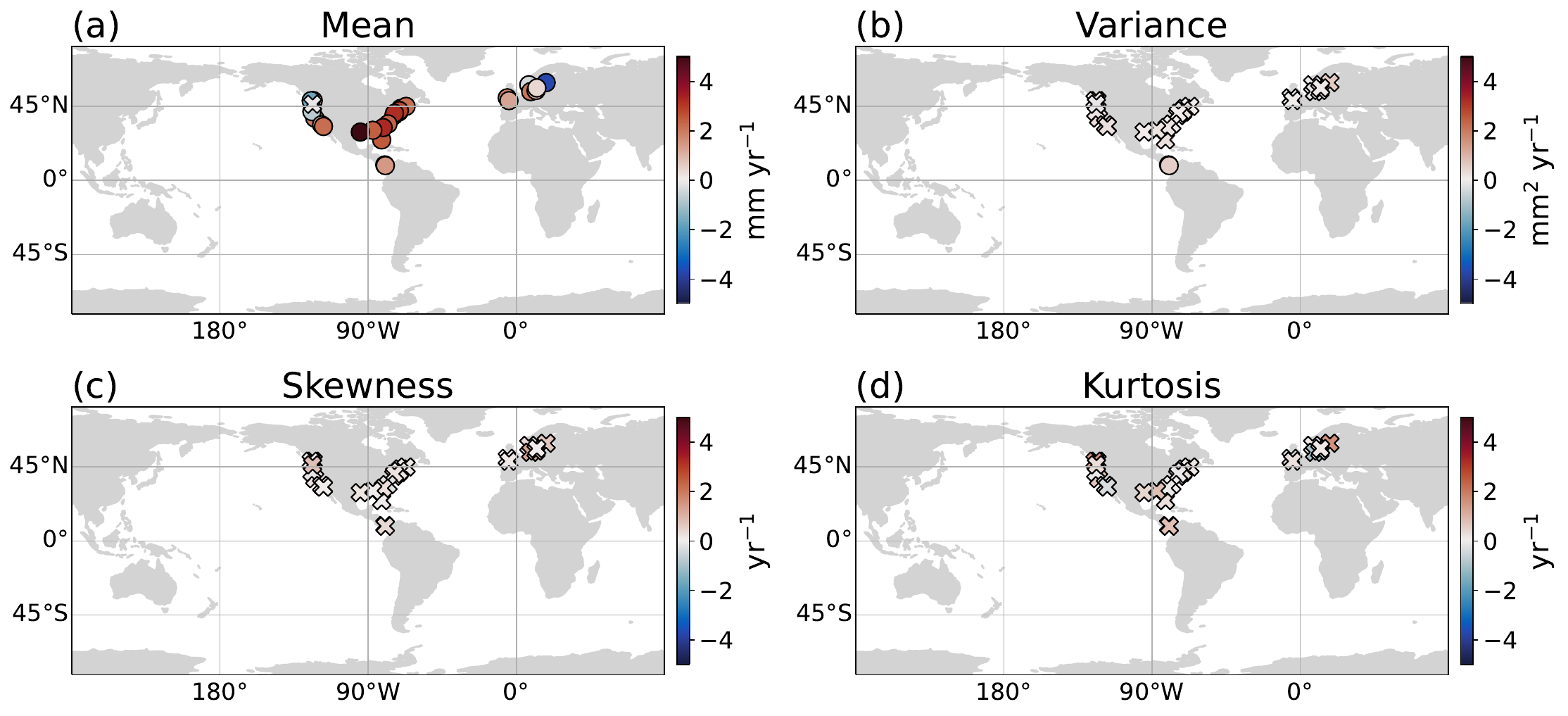}
\caption{Projection onto mean, variance, skewness and kurtosis functions. All tide gauges considered have a record longer than 80 years and less than 20$\%$ of missing data. Statistically significant changes are marked with filled ``circles'', whereas insignificant changes are marked with ``X''s. Statistical significance is computed using the FDR test with threshold $\phi = 0.05$ for the mean, variance, skewness and kurtosis slopes. The starting and ending date for each tide gauge is shown in the Appendix, Section \ref{app:long_records_tide_gauges}}  
\label{fig:fdr_tide_gauges_long_record}
\end{figure}

We further analyze tide gauges with (a) more than 80 years of data and (b) less than 20$\%$ of missing values, with results presented in Figure \ref{fig:fdr_tide_gauges_long_record} and Table \ref{table:significance_obs}. The record spanned in such locations is reported in the Appendix, Section \ref{app:long_records_tide_gauges}. The significance test adopted is FDR. Also in this case we observe significant changes in the distributional mean but not in higher order moments. As an exception, two records from Panama (i.e., Cristobal and Balboa) show a significant change in variance. Such changes can be partially explained by very large anomalous El Ni\~no and La Ni\~na events since 1980, as shown in the Appendix, Section \ref{app:p-values-tide-gaues}.\\

We conclude that, independently of the period analyzed, changes in tide gauge measured coastal sea level distributions can be characterized by a shift in the mean with no statistically significant change in shape. 

\begin{table}
\caption{Percentage of significant coefficients for the FDR test. All figures report results obtained with the FDR test with a threshold $\phi = 0.05$ (see Section \ref{sec:significance_test}).}
\centering
\begin{tabular}{l c c c c c}
\hline
  & Mean & Variance & Skewness & Kurtosis &   \\
\hline
\textcolor{blue}{Period 1970-2017}\\
FDR ($\phi = 0.05$)  & $92$($97.9\%$) & $4$($4.3\%$) & $1$($2.1\%$) & $0$($0.0\%$) & \small{out of 94}   \\
\hline
\textcolor{blue}{Records $\geq$ 80 years}\\
FDR ($\phi = 0.05$)  & $27$($96.4\%$) & $2$($7.1\%$) & $0$($0.0\%$) & $0$($0.0\%$) & \small{out of 28}   \\
\hline
\end{tabular}
\label{table:significance_obs}
\end{table}

\clearpage

\subsection{GFDL-CM4 climate model}

\begin{figure}[tbhp]
\centering
\includegraphics[width=1\linewidth]{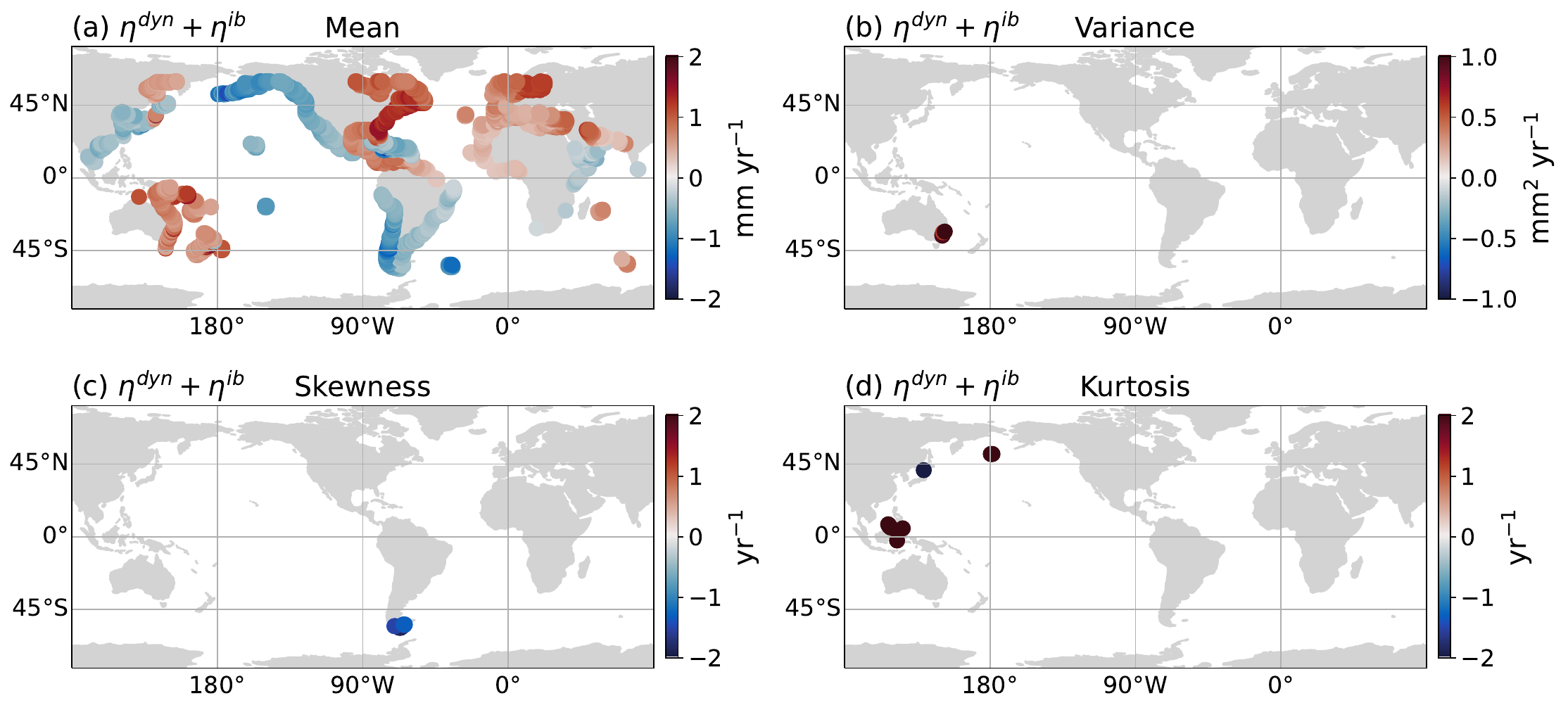}
\caption{Projection onto mean, variance, skewness and kurtosis functions for the GFDL-CM4 historical experiment. We consider the dynamic sea level and inverse barometer (i.e., $\eta^{\mbox{\tiny dyn}} + \eta^{\mbox{\tiny ib}}$) contributions in the period 1970-2014. Statistical significance is computed using the FDR test with threshold $\phi = 0.05$, respectively, for the mean, variance, skewness and kurtosis slopes. The global thermosteric contribution is not included in the analysis. Only statistical significant slopes are reported}
\label{fig:slopes_model_historical}
\end{figure}

We quantify changes in sea level distributions in the historical experiment with CM4, focusing on the dynamic sea level plus inverse barometer (i.e., $\eta^{\mbox{\tiny dyn}} + \eta^{\mbox{\tiny ib}}$) during the period 1970-2014. Results are shown in Figure \ref{fig:slopes_model_historical} and significant changes are reported in Table \ref{table:significance_model}. In this period, roughly $43\%$ of the coastal area experiences significant changes in the mean sea level. Among such changes, positive sea level rise is detected mainly along the US east coast, North and East Africa, as well as Europe and Oceania. Negative trends are simulated along coastlines in East Asia, western North America, and almost the whole South American coast. Importantly, GFDL-CM4 does not exhibit changes in the shapes of distributions in the historical period, which agrees with our tide gauge analysis (Table \ref{table:significance_model}).\\

\begin{figure}
\centering
\includegraphics[width=1\linewidth]{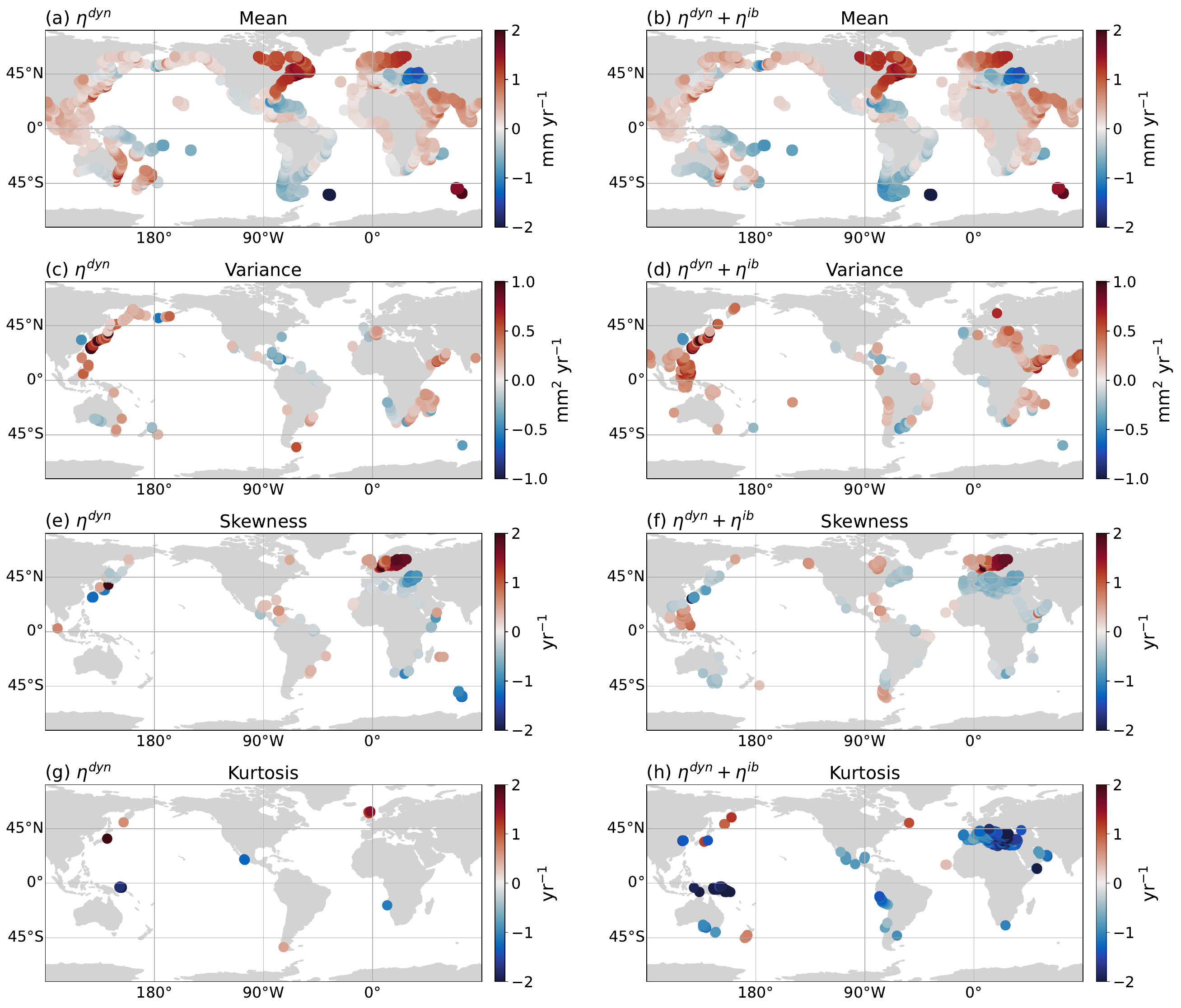}
\caption{Projection onto mean, variance, skewness and kurtosis functions for the GFDL-CM4 experiment with 1$\%$ $\text{CO}_{2}$ yearly increase for 100 years. Panels (a,c,e,g): dynamic sea level only (i.e., $\eta^{\mbox{\tiny dyn}}$). Panels (b,d,f,h): dynamic sea level and inverse barometer contributions (i.e., $\eta^{\mbox{\tiny dyn}} + \eta^{\mbox{\tiny ib}}$). Statistical significance is computed using the FDR test with threshold $\phi = 0.05$ respectively for the mean, variance, skewness and kurtosis slopes. The global thermosteric contribution is not included in the analysis. Only statistical significant slopes are reported}
\label{fig:slopes_model_1pctCO2}
\end{figure}

We next explore changes in statistics in the last 100 years of the 1pctCO2 experiment with results given in Figure \ref{fig:slopes_model_1pctCO2} and Table \ref{table:significance_model}. To investigate the main physical drivers of distributional shifts, we first analyze the dynamic sea level $\eta^{\mbox{\tiny dyn}}$ (Figure \ref{fig:slopes_model_1pctCO2}, left column) and then consider the effect of the inverse barometer, i.e., $\eta = \eta^{\mbox{\tiny dyn}} + \eta^{\mbox{\tiny ib}}$ (Figure \ref{fig:slopes_model_1pctCO2}, right column). We observe that in both cases a large number of coastal locations experience a significant change in the mean of the distribution (Figure \ref{fig:slopes_model_1pctCO2}(a,b)). The fraction of significant changes is slightly lower when the inverse barometer effect is included ($\sim67\%$ vs $\sim79\%$) but still much larger than the $43\%$ found in  the historical experiment. One difference with the historical experiment is the emergence of mean sea level changes in the South and South-East part of Africa, Indonesia and India. Additionally, many negative trends in mean sea level found in the historical experiment switch sign, especially in the North-West America, East Asia (i.e., Sea of Japan, East and South China sea) and East Africa.\\

A major difference with both the observational dataset (Section \ref{sec:all_tide_gauges}) and the CM4 historical experiment is the emergence of changes in higher order moments in the 1pctCO2 experiment. This emergence shows that changes in shapes of daily sea level distributions are indeed possible, at least in simulations as $\text{CO}_{2}$ increases. We detect coherent positive significant changes in variance along the Japan coast (Figure \ref{fig:slopes_model_1pctCO2}(c,d)). In addition, when also the inverse barometer  is included, we see the emergence of positive changes in variance in part of Indonesia and in North- and South-East Africa and East India (Figure \ref{fig:slopes_model_1pctCO2}d). Adding the inverse barometer component always implies a larger fraction of significant changes in second, third and fourth moments. This result holds for shifts in variance and it is much more evident in the case of third and fourth moments as shown in Table \ref{table:significance_model}. Spatially coherent changes in skewness include positive trends around North Indonesia and in the Philippines and Northern Europe (Figure \ref{fig:slopes_model_1pctCO2}f). Interestingly, we also detect many significant negative changes in skewness, mainly in Japan coasts, North East US, virtually all coasts in the Mediterranean Sea and North East Africa (Figure \ref{fig:slopes_model_1pctCO2}f).\\

Future work will further address possible causes of such higher order changes. Hereafter we focus mainly on drivers of shifts in kurtosis and on large, spatially coherent changes in the Mediterranean sea and in Northern Europe.\\

Shifts in the kurtosis functions are negligible in the dynamic sea level contributions and detected only when the inverse barometer is included (Table \ref{table:significance_model}). The largest, spatially coherent domain with changes in kurtosis is found in the Mediterranean, as shown in Figure \ref{fig:slopes_model_1pctCO2}h. This area is marked by significant (negative) changes in both skewness and kurtosis functions. To further explore such distributional shifts, we considered the time series with largest changes in kurtosis in the Mediterranean and analyzed it separately. Results are shown in the Appendix, Section \ref{app:Med-sea}. The histograms for the first and last 40 years of data (see Figure S13(g,h)) show a decrease in skewness from $-0.06$ to $-0.37$ and an excess kurtosis from $0.83$ to $0.03$ (closer to the Gaussian case of $0$).\\

Such changes in skewness and kurtosis in the Mediterranean come \textit{solely} from changes in sea level pressure (SLP) anomalies (see Figure \ref{fig:slopes_model_1pctCO2}(g,h)). This result indicates a drop in frequency of (large) negative SLP anomalies (see Eq. \ref{eq:sea_level_decomposition}) and points to a large decrease in frequency of hurricanes in the Mediterranean, the so-called ``medicanes'', as recently suggested in \cite{Aleman}. We note that while the frequency of medicanes is projected to decrease, the strongest medicanes are expected to intensify \citep{Aleman}.\\

Changes in kurtosis are present in regions outside the Mediterranean sea. The majority of such changes are (i) found to be negative and (ii) coming only from the inverse barometer effect (Figure \ref{fig:slopes_model_1pctCO2}(h)). This possibly points out to changes in the statistics of intense convective systems. This seems to be large in agreement with \cite{Priestley} where the authors quantified a robust decrease in cyclone numbers, independent of the season, in CMIP6 models projections. On the other hand, the strongest cyclones are projected to have higher intensities (mean sea level pressure and vorticity) and larger tropospheric wind speed (with changes dependent on different seasons and hemispheres) \citep{Priestley}.\\

A large number of coastal location experience positive changes in skewness in Northern Europe (North sea and Baltic sea). Such changes are already present in the dynamic sea level field, $\eta^{\mbox{\tiny dyn}}$, and show no qualitative changes when the inverse barometer is included. Positive changes in skewness in the dynamic sea level field possibly indicate a shift to larger frequency of storm surges. This is in agreement with \cite{Gaslikova}, where the authors studied changes in North sea storm surges in future climate scenarios (i.e., 1961–2100 period) by comparing changes in the 99th percentiles of water levels in 30-year windows. \cite{Gaslikova} used a hydrodynamical model and quantified a (small) increase in frequency of such extreme events, consistent with an increase in frequency of intense westerly winds. An increase in frequency of positive large sea level rise anomalies (positive slopes in skewness) in that region is then consistent with such increase in wind forcing. Importantly, such anthropogenic signals are superimposed on large, decadal oscillations leading to uncertainties coming from the system's internal variability.\\

Spatially coherent changes in extreme sea levels over the Mediterranean sea and Northern Europe can be in part linked to projected changes in atmospheric storm track activity and cyclone intensities over Europe as shown by \cite{pinto}. The authors analyzed ensembles of climate change projections and identified particularly strong reductions of cyclone intensities in subtropical areas such as in the Mediterranean sea and a large increase in extreme surface winds in Great Britain, North and Baltic seas. Importantly, the role of internal variability was shown to be important and large differences were identified among ensemble members \citep{pinto}. Such changes were found to mostly hold in the new generation of climate models, and \cite{Priestley} quantified a decrease in cyclone activity over the Mediterranean in CMIP6 future projections, together with changes in the North Atlantic storm track.\\

\begin{table}
\caption{Percentage of significant coefficients for GFDL-CM4 historical and 1pctCO2 experiments. For the historical experiment we consider the dynamic sea level and inverse barometer (i.e., $\eta^{\mbox{\tiny dyn}} + \eta^{\mbox{\tiny ib}}$) contributions in the period 1970-2014.  For the 1pctCO2 experiment we show results both for the dynamic sea level and inverse barometer (i.e., $\eta^{\mbox{\tiny dyn}} + \eta^{\mbox{\tiny ib}}$) and dynamic sea level only (i.e., $\eta^{\mbox{\tiny dyn}}$). Statistical significance is computed using the FDR test with threshold $\phi = 0.05$.}
\centering
\renewcommand{\arraystretch}{1.3}
\begin{tabular}{l c c c c c}
\hline
 & Mean & Variance & Skewness & Kurtosis &  \\
\hline
\textcolor{blue}{Period 1970-2014}\\
$[\eta^{\mbox{\tiny dyn}} + \eta^{\mbox{\tiny ib}}]$; FDR ($\phi = 0.05$)  & $2702$($42.8\%$) & $5$($0.1\%$) & $11$($0.2\%$) & $11$($0.2\%$) & \small{out of }6318   \\
\hline
\textcolor{blue}{1$\%$ $\text{CO}_{2}$ yearly increase}\\
$[\eta^{\mbox{\tiny dyn}}]$; FDR ($\phi = 0.05$)  & $4939$($78.8\%$) & $293$($4.6\%$) & $315$($5.0\%$) & $18$($0.3\%$) & \small{out of }6318   \\
$[\eta^{\mbox{\tiny dyn}} + \eta^{\mbox{\tiny ib}}]$; FDR ($\phi = 0.05$)  & $4260$($67.4\%$) & $632$($10.0\%$) & $914$($14.5\%$) & $394$($6.2\%$) & \small{out of }6318   \\
\hline
\end{tabular}
\label{table:significance_model}
\end{table}

\clearpage

\section{Conclusions} \label{sec:conclusions_discussions}

An in-depth assessment of changes in shapes of sea level distributions in observations and in climate simulations has not been addressed in the current literature. In fact, a large focus in the last Intergovernmental Panel on Climate Change (IPCC) report has been on changes in the mean of the distributions \citep{ipccChapter9}. Furthermore, changes in sea level extremes through extreme value theory are often quantified solely as a function of changes in the distributional mean \citep{Tebaldi}. This motivated us to address the following questions regarding changes in sea level probability distributions: 
\begin{itemize}
    \item Are changes in higher order moments in observations (a) significant but small compared to the shift in the mean or (b) consistent with the climate system's internal variability? 
    \item How do shapes of sea level distribution evolve under $\text{CO}_2$ forcing at centennial time scales, at least in a climate model framework?  Which sea level component (i.e., sterodynamic and inverse barometer effects) is responsible for such changes?
\end{itemize}
Our work addressed and answered these questions from a novel statistical point of view by exploring changes in sea level distributions in tide gauges and in the GFDL-CM4 model.\\

To achieve these goals, we introduced a novel and general statistical framework to study changes in both quantiles and moments of a distribution and quantify their significance under the system's internal variability. Importantly, shifts in moments are captured by suitable \textit{orthogonal} functions, therefore capturing \textit{independent} sources of distributional changes. The framework is inspired by the methodology proposed in \cite{mckinnon2016changing} but differs in terms of polynomials and statistical tests. Here the choice of polynomials follows from the rigorous theoretical work of \cite{Cornish} and \cite{Fisher}. Their theory allows us to derive a precise formalism to investigate changes in both quantiles and moments. Additionally, the recent work of \cite{Chernozhukov} allows for very fast computation of quantile regression, thus making the methodology feasible even when dealing with thousands of time series each with tens of thousands of time steps, as commonly found with climate model applications. In the case of a drifting Gaussian distribution, the first and second polynomials quantify the \textit{exact} link between changes in moments and quantiles. The third and fourth polynomials adjust for non-Gaussian behaviour. Future work may consider the Cornish-Fisher expansion for asymptotic estimates of very large quantiles otherwise difficult (or impossible) to do from data alone, therefore providing a useful alternative to extreme value theory.\\

The statistical significance test adopts the False Discovery Rate proposed by \cite{Benjamini}, therefore controlling against multiple testing. The bootstrapping methodology is robust under the sample size $B=1000$. In general, even for sample sizes as small as $B = 100$ we can still obtain meaningful results, as shown for one time series in Appendix \ref{app:balboa_record}, Figure \ref{fig:bootstrap_tests}. However, as in any other resampling test, a larger number of samples (i.e., $B=5000$) would probably be preferred and here avoided because of long run time (same as in \cite{mckinnon2016changing}).  The methodology is linear but still relevant in the case of weak nonlinearities, with an example shown for the Beta distribution in Section \ref{subsec:synthetic}. Strongly nonlinear trends in the mean of the distribution will not be detected by this linear framework, but such nonlinearities would result in significant changes in higher order moments. The underlying physical mechanisms leading to changes in sea level distributions can be nonlinear, such as for a weakening Atlantic meridional overturning circulation (AMOC) commonly found in climate model projections \citep{Weijer}. For example, in the GFDL-CM4 climate model simulation considered here, the AMOC strength weakened by roughly a factor of two after 100 years of enhanced CO2 forcing (see Figure 13b of \cite{jianjun}). Even so, statistical properties of the sea level distribution retained a close to linear behavior, thus allowing for the methodology developed in this paper to remain useful.\\

We applied the above statistical  framework to coastal daily sea level measured by tide gauges. Nearly every record shows a significant change in the mean of the distribution for the period 1970-2017. In contrast, we detect no significant change in higher order moments for the same period, with this conclusion robust to sampling only tide gauges with more than 80 years of data. We conclude that changes in coastal sea level in the historical period exhibited just a shift in the mean, with no significant change in the shape of the probability distribution.  Likewise, no changes in higher order statistical moments are found in the GFDL-CM4 historical simulation. Hence, changes in the simulated coastal sea level arise mainly by changes in the mean, which agrees with the tide gauges. We note that multidecadal oscillations could in principle impact statistical attribution in some locations when focusing on ``short'' datasets. However, the same conclusions are obtained for tide gauges with 47 years (period 1970-2017) and more than 80 years of data. Conclusions on observations are therefore robust on the chosen period.\\

We then considered a CM4 experiment with 1$\%$/yr $\text{CO}_{2}$ forcing, thus allowing us to quantify changes in sea level in a warming (modeled) world. In this experiment, we find (i) a large increase in significant changes in the mean of the distribution compared to the historical run and (ii) the emergence of changes in higher order moments.\\

Interestingly, changes in the second and third moments are already present in the dynamic-sea-level-only analysis. These changes are therefore driven solely by the evolution of ocean circulation, caused by changes in water column mass and local steric effects. Such changes get amplified when sea level pressure fluctuations are included, accounting for the inverse barometer effect \citep{Ponte}. Shifts in kurtosis are, for the most part negative, and possibly consistent with results of \cite{Priestley} where the authors observed robust decrease in cyclone numbers, independent of the season, in CMIP6 models projections. One striking difference with the dynamic-sea-level-only analysis is the emergence of large changes in skewness and kurtosis along the Mediterranean coast, possibly pointing to drops in the frequency of medicanes, as shown in \cite{Aleman}. A decrease in frequency of medicanes has been shown to hold on average for different regional models, but some model runs show no significant changes \citep{ROMERA2017134}. Larger ensembles and different scenarios are then needed to better quantify the impact of statistical changes in medicanes on coastal sea level. Finally, a large increase in skewness values in the North and Baltic seas is present in the dynamic-sea-level-only analysis and show no (qualitative) differences with the inclusion of inverse barometer. This is consistent with an increase in frequency of intense westerly winds in that region as shown by \cite{pinto} and \cite{Gaslikova}.\\

We note that the GFDL-CM4 does not simulate category 4 and 5 tropical cyclones (see \cite{jianjun}). Hence, shifts in higher order moments of the distributions are likely underestimated in the simulation. Such changes in cyclones frequency should be further explored using different models and scenarios. Furthermore, such information should be taken into account in the context of extreme value attribution where extreme sea level is often modeled solely as a function of the distributional mean.\\

A limitation of this work is that we focused only on one climate model and we did not consider ensembles of simulations. First, the climate is a chaotic system and ensemble simulations are often needed to explore the system's variability. Second, climate models are affected by structural model errors and a collection of models is required for more confident climate projections \citep{Frigg,LennyPNAS}. The emergence of significant changes in higher order moments in the model analyzed here should be seen as a \textit{proof of concept} rather than a quantitative prediction about the \textit{real} world. However, this result clearly suggests that quantifying sea level changes solely as a function of the mean of the distribution can be too simplistic. Future work should focus on different models, ensemble runs and scenarios in order to quantify possible uncertainties coming from different model structures and internal variability.\\

Results of this study could guide new analyses focused on regions experiencing changes in higher order moments. An example would be to make use of higher resolution regional models to further study dynamical processes in those locations. Additionally, future studies could focus on changes in high (i.e., 0.95) and small (i.e., 0.05) quantiles only in such locations across different models and scenarios. Finally, future work will explore changes in the probability distribution of relevant climate variables other than sea level, from temperature to precipitation, therefore contributing to a better understanding of climate variability and its linkages with natural and anthropogenic forcing.\\

\appendix

\section{Preprocessing tide gauge data} \label{app:preprocessing}

Quality-controlled sea level research-grade tide gauge data were downloaded from the University of Hawaii sea level center \cite{Caldwell} (\url{https://uhslc.soest.hawaii.edu/}).\\

Few locations are associated with multiple distinct tide gauges (indicated by different record IDs). This results in two main issues: (a) different tide gauge records are often measured with respect to different reference levels; (b) some tide gauges have temporally overlapping records. Ideally, such different records should be concatenated into a single timeseries in a way that remains faithful to the trends observed at each individual station. We address these two issues by implementing the following pipeline for preprocessing the data:
\begin{itemize}
    \item linear trends from each record are computed.
    \item A residual timeseries is obtained by detrending each record and concatenating the timeseries. If any part of two or more records overlap temporally, those portions of the residual timeseries are averaged.
    \item Slopes for the data are defined at each time by computing the average trend over each record ID. Gaps in the slopes are linearly interpolated. A composite trendline is created by cumulatively integrating the slopes over time.
    \item The residual timeseries is added to the composite trendline to obtain a composite timeseries.
\end{itemize}

\begin{figure}
    \centering
    \includegraphics[width=0.8\linewidth]{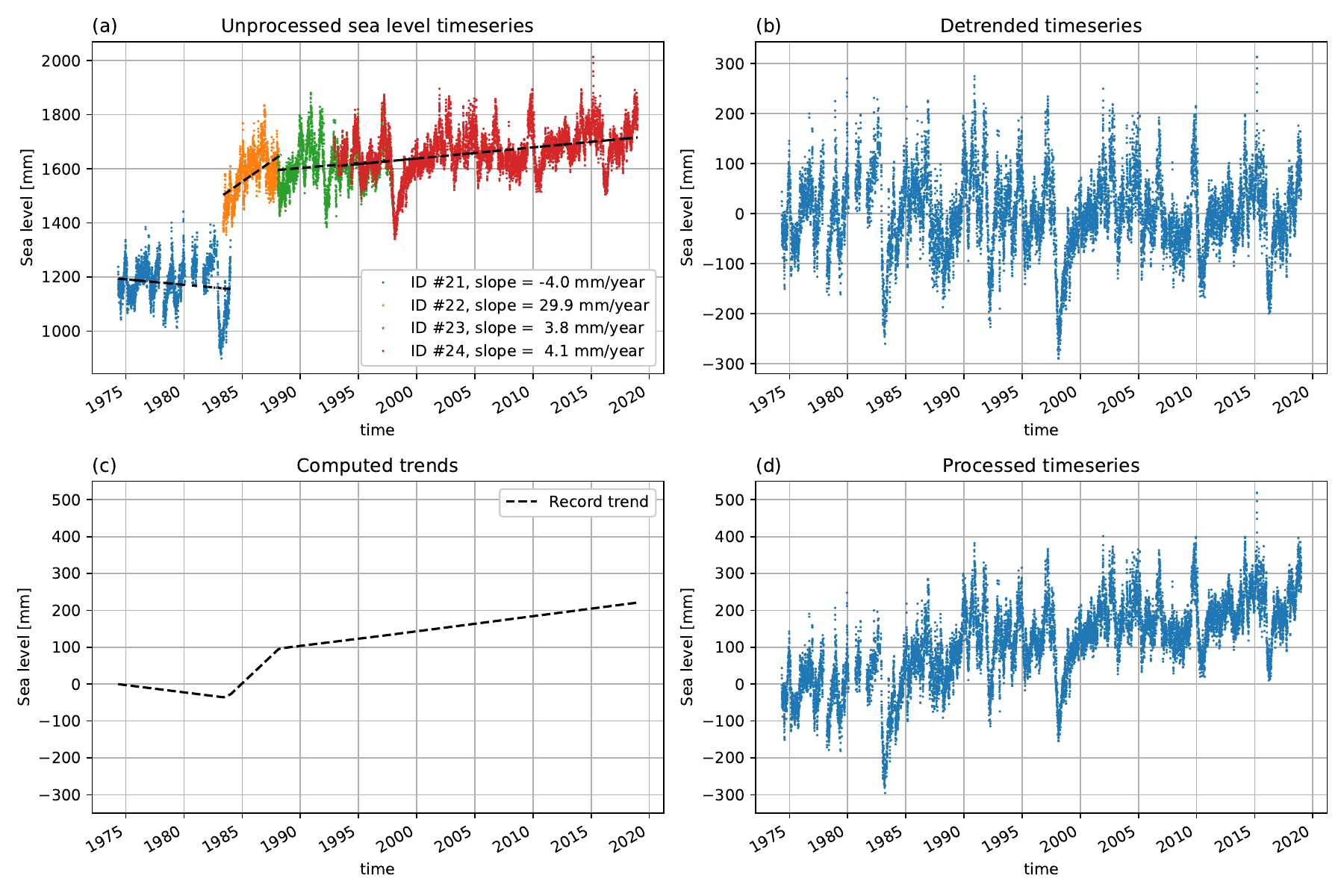}
    \caption{Illustration of preprocessing of tide gauge for observations at Tarawa, Kiribati. {(a)} Original timeseries, with different records demarcated by different colors. Note the different reference points and overlapping timespans. {(b)} Residual (detrended) timeseries obtained by detrending records and concatenating values, averaging observations over different records if necessary. {(c)} Computed trendline obtained by integrating record-averaged slopes over time. {(d)} Composite processed timeseries obtained by adding (b) and (c)}
    \label{fig:tide-gauge-processing}
\end{figure}

Figure~\ref{fig:tide-gauge-processing} illustrates the process for a single tide gauge record at Tarawa, Kiribati.

\clearpage

\section{Tide gauges with long records} \label{app:long_records_tide_gauges}

In Table \ref{table:start_end_dates} we show the start and end dates for each one of the long records tide gauges considered. Each one of these records include more than 80 years of data and has less than 20$\%$ missing data.

\begin{table}
\caption{Location, starting and ending date of tide gauges with (a) records longer than 80 years and (b) less than 20$\%$ missing data.}
\renewcommand{\arraystretch}{1.}
\begin{tabular}{l c c}
\hline
Tide Gauge & Start date & End date  \\
\hline
Key West (FL) & 1913-01-20 & 2018-12-31 \\
Portland (ME) & 1910-03-05 & 2018-12-31 \\
Newport (RI) & 1930-09-11 & 2018-12-31 \\
Atlantic City (NJ) & 1911-08-20 & 2018-12-31 \\
Cristobal (Panama) & 1907-04-04 & 2017-12-31 \\
Newlyn (Cornwall) & 1915-04-23 & 2016-12-31 \\
Balboa (Panama City) & 1907-06-20 & 2018-12-31 \\
Victoria (BC) & 1909-02-19 & 2018-12-31 \\
San Francisco (CA) & 1897-08-02 & 2018-12-31 \\
La Jolla (CA) & 1924-10-03 & 2018-12-31 \\
Crescent City (CA) & 1933-04-12 & 2018-12-31 \\
Neah Bay (WA) & 1934-08-02 & 2018-12-31 \\
Los Angeles (CA) & 1923-11-29 & 2018-12-31 \\
San Diego (CA) & 1906-01-22 & 2018-12-31 \\
Astoria (OR) & 1925-01-26 & 2018-12-31 \\
Eastport (ME) & 1929-09-13 & 2018-12-31 \\
Boston (MA) & 1921-05-04 & 2018-12-31 \\
New York (NY) & 1920-06-02 & 2018-12-31 \\
Wilmington (NC) & 1935-12-29 & 2018-12-31 \\
Fort Pulaski (GA) & 1935-07-02 & 2018-12-31 \\
Pensacola (FL) & 1923-05-02 & 2018-12-31 \\
Galveston Pier 21 (TX) & 1904-01-02 & 2018-12-31 \\
Tregde (Norway) & 1927-10-05 & 2018-12-31 \\
Brest (France) & 1846-01-04 & 2018-12-31 \\
Cuxhaven (Germany) & 1917-12-30 & 2018-12-31 \\
Stockholm (Sweden) & 1889-01-01 & 2014-12-31 \\
Gedser (Denmark) & 1891-09-02 & 2012-12-31 \\
Hornbaek (Denmark) & 1891-01-02 & 2012-12-31 \\
\end{tabular}
\label{table:start_end_dates}
\end{table}

\clearpage

\section{\textit{Exact} and \textit{sample} moments of a distribution} \label{app:moments_definition}

Consider a probability density function $P(x)$, with $x \in \mathbb{R}$. $P(x)$ satisfies $\int_{-\infty}^{\infty} P(x) dx = 1$. The first moment, i.e. the mean, is defined as:

\begin{linenomath}
\begin{equation}
\mu = \int_{-\infty}^{\infty} x P(x) dx.
\label{eq:mean}
\end{equation}
\end{linenomath}

The central moment $\mu_{n}$ of order $n$ is defined as the $n$-th moment about the mean as:

\begin{linenomath}
\begin{equation}
\mu_{n} = \int_{-\infty}^{\infty} (x - \mu)^{n} P(x) dx.
\label{eq:central_moment}
\end{equation}
\end{linenomath}

The variance $\sigma^{2}$ of the distribution is defined as the central moment of order $n = 2$, i.e. $\mu_{2}$:

\begin{linenomath}
\begin{equation}
\sigma^{2} = \mu_{2} = \int_{-\infty}^{\infty} (x - \mu)^{2} P(x) dx.
\label{eq:variance}
\end{equation}
\end{linenomath}

The skewness $\gamma$ and excess kurtosis $\kappa$ of the distribution are defined as the 3rd and 4th standardized central moments:

\begin{linenomath}
\begin{align}
\gamma &= \frac{\mu_{3}}{\mu_{2}^{3/2}};\\
\kappa &= \frac{\mu_{4}}{\mu_{2}^{2}} - 3.
\label{eq:skewness_and_kurtosis}
\end{align}
\end{linenomath}

Skewness and kurtosis are invariant under translation and rescaling and therefore capture exclusively the shape of the distribution. The focus on excess kurtosis is only a matter of practical convenience as the kurtosis of a normal distribution is equal to 3. In this paper we are interested in the temporal evolution of the mean ($\mu$), variance ($\sigma^{2}$), skewness ($\gamma$), and excess kurtosis ($\kappa$).\\

Note that when focusing on data we can only compute the \textit{sample} mean, variance, skewness and kurtosis which serve only as estimators of the \textit{exact} quantities as defined above. Given a set of $N$ observations ${x_{1},x_{2},...,x_{N}}$, we define the \textit{sample} mean as: 

\begin{linenomath}
\begin{equation}
\hat{\mu} = \frac{1}{N}\sum_{i=1}^{N}x_{i}.
\label{eq:sample_mean}
\end{equation}
\end{linenomath}

The \textit{sample} central moment of order $n$ is then defined as 

\begin{linenomath}
\begin{equation}
\hat{\mu}_{n} = \frac{1}{N-1}\sum_{i=1}^{N}(x_{i}-\hat{\mu})^{n}.
\label{eq:sample_central_moment}
\end{equation}
\end{linenomath}

Therefore, the \textit{sample} variance ($\hat{\sigma}^{2}$), skewness ($\hat{\gamma}$) and excess kurtosis ($\hat{\kappa}$) are defined as:

\begin{linenomath}
\begin{align}
\hat{\sigma}^{2} &= \frac{1}{N-1}\sum_{i=1}^{n}(x_{i}-\hat{\mu})^{2}\\
\hat{\gamma} &= \frac{\hat{\mu}_{3}}{\hat{\mu}_{3}^{3/2}}\\
\hat{\kappa} &= \frac{\hat{\mu}_{4}}{\hat{\mu}_{4}^{2}} - 3.
\label{eq:sample_higher_order_moments}
\end{align}
\end{linenomath}

In the main text we often refer to ``moments'' without specifying if they are \textit{sample} or \textit{exact} moments. The fitting procedure derived in Section \ref{subsubsec:derivation} aims in finding changes in the \textit{exact} statistical moments. The Cornish-Fisher tests shown in Sections \ref{app:beta_test} and \ref{app:CF_sea_level} make use of the \textit{exact} and \textit{sample} moments respectively.\\

Importantly, for simplicity and to ease the derivation in Section \ref{subsubsec:derivation}, all throughout the paper we refer to the mean, variance, skewness and kurtosis as $(m_{1},m_{2},m_{3},m_{4})$.

\section{Cornish-Fisher Expansion} \label{app:CF_testing}

The Cornish-Fisher expansion allows for asymptotic estimations of quantiles of a probability distribution in terms of its first four moments \citep{Cornish}. We refer the reader to Section \ref{subsubsec:Cornish-Fisher} for more details on such formula. Estimations are valid for distributions that do not show large deviations from normality (i.e., reltively small skewness and kurtosis) with \cite{Maillard} showing the domain of validity of such formula. \cite{Maillard,Manesme} and \cite{Lamb} showed how to correct the skewness and kurtosis parameters in the case of larger deviation from normality. These corrections are not necessary in our study where the polynomials in Section \ref{subsubsec:derivation} are defined in the limit of $m_{3}$ and $m_{4}$ approaching zero. In this case $m_{3}$ and $m_{4}$ are \textit{exactly} the skewness and kurtosis of the distribution.\\ 

Additionally, here we present two tests using the original expression shown in Section \ref{subsubsec:Cornish-Fisher}. In the first test we show how the Cornish-Fisher expansion allows for a trustworthy approximation of the quantile function of a Beta distribution. In a second step, we extend the analysis for sea level stationary timeseries and show remarkable agreement with the estimated quantiles as computed using the NumPy quantile function \citep{harris2020array}. Future work may focus on quantile estimations using the corrected versions such as in \cite{Maillard,Manesme,Lamb} to account for large deviations from normality (even if this may not often be necessary for climate observables). Importantly, we note that for large values of skewness and kurtosis we expect large errors for very small and very large quantiles (e.g., $q_{p}$ with $p = 0.0001$ or $p = 1 - 0.0001$) but relatively smaller errors for less extreme values, such as the one considered in this paper (e.g. $q_{p}$ with $p\in[0.05,0.95]$).

\subsection{Test on Beta distribution}\label{app:beta_test}

\begin{figure}[tbhp]
\centering
\includegraphics[width=1\linewidth]{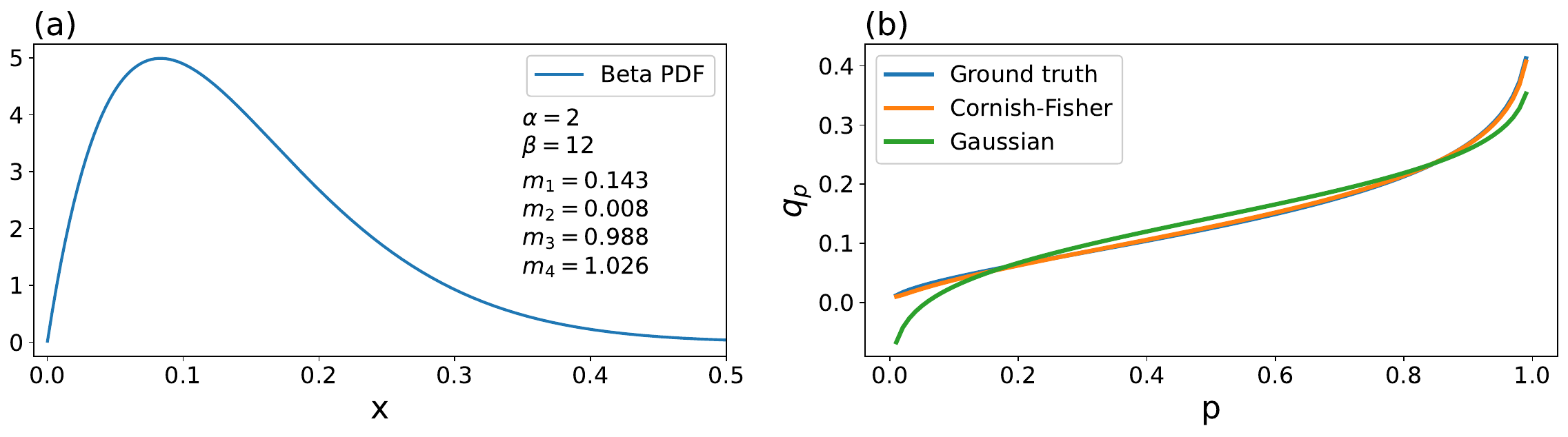}
\caption{A test for the Cornish-Fisher expansion in the case of a Beta distribution. Panel (a): probability density function of a beta distribution with $\alpha = 2$ and $\beta = 12$. Formulas defining a Beta distributions are shown in Section \ref{subsec:synthetic}. The mean ($m_{1}$), variance ($m_{2}$), skewness ($m_{3}$) and excess kurtosis ($m_{4}$) are also reported. Panel (b): estimation of quantiles $q_{p}$ with $p\in[0.01,0.99]$ every $dp=0.01$. In blue we show the ground truth which can be computed analytically for the Beta distribution. In green we show the estimation obtained under a Gaussian assumption; in this case, the mean $m_{1}$ and variance $m_{2}$ are enough to compute any quantile. In orange we show the Cornish-Fisher estimation allowing to correct for non-normality by computing quantiles as a function of the first 4 moments}
\label{fig:beta_CF_test}
\end{figure}

We consider a Beta distribution (see Section \ref{subsec:synthetic}) with parameters $\alpha = 2$ and $\beta = 12$ and show it in Figure \ref{fig:beta_CF_test}(a). In this case the exact (not sample) mean ($m_{1}$), variance ($m_{2}$), skewness ($m_{3}$) and excess kurtosis ($m_{4}$) take the values $(m_{1},m_{2},m_{3},m_{4}) = (0.143,0.008,0.988,1.026)$.The probability distribution is defined for $x\in[0,1]$.\\

Given this Beta distribution, we aim in computing quantiles $q_{p}$ with $p\in[0.01,0.99]$ every $dp = 0.01$ in three different ways.

\begin{itemize}
    \item We compute the exact quantiles of the Beta distribution with $\alpha = 2$ and $\beta = 12$; see blue curve in Figure \ref{fig:beta_CF_test}(b).
    \item We estimate quantiles under the (wrong) assumption of Gaussianity (green curve in Figure \ref{fig:beta_CF_test}(b)). In this case, every quantile can be estimated by the mean and variance of the distribution.
    \item We correct the previous assumption of normality through the Cornish-Fisher expansion. This allows for asymptotic estimations of the quantiles of the distributions as function of the first four moments. Differences from the ground truth estimation are very small (orange curve in Figure \ref{fig:beta_CF_test}(b)).
\end{itemize}

\subsection{A sea level test}\label{app:CF_sea_level}

The Cornish-Fisher expansion allows for the computation of any quantile of general distributions as deviations from the corresponding quantile of a normal distribution \cite{Cornish,Fisher}. As far as we know, this approximation has not been used in climate studies. Here we explore its relevance for sea level studies.\\

We consider the last 300 years of a 650 years long piControl experiment simulated with the GFDL-CM4 model. The model's specifics are described in Section \ref{sec:data}. The piControl experiment is forced by constant $\text{CO}_{2}$ forcing, set at pre-industrial level. Hence, time series at each grid point are close to stationary, with climate model drift relatively small for the 300 years analyzed here. In order to limit the amount of computations in this section, we remapped the ocean model grid to a uniform 1$^\circ$ grid and only points in the latitudinal range $[-60^\circ,60^\circ]$ are considered. Furthermore, we consider 3-day averages, therefore accounting for 36500 time steps at each grid point. The variable of interest is the dynamic sea level (i.e., ``zos'', see Section \ref{eq:sea_level_decomposition}), hereafter referred to as $\eta^{\mbox{\tiny dyn}}$ for consistency with the main discussion in Section \ref{sec:results}.\\

For a sea level time series $\eta^{\mbox{\tiny dyn}}(x,y,t)$ at longitude and latitude $(x,y)$, we compute the 0.95 quantile, $q_{0.95}$. We do so in three different ways.

\begin{itemize}
    \item We compute $q_{0.95}$ using the quantile function in the Python, NumPy package \citep{harris2020array}. We refer to this value as ``Ground Truth'' or ``GT'' under the assumption that 300 years long (stationary) time series are enough to have a good estimate of the true, underlying distribution.
    \item We assume that each time series follows a Gaussian distribution. In this case the 0.95 quantile of each $\eta^{\mbox{\tiny dyn}}(x,y,t)$ can be computed \textit{exactly} by knowing its mean $m_1$ and variance $m_2$.
    \item We adopt the Cornish-Fisher expansion to extend the previous computation to non-normal statistics. Therefore, we ``correct'' the Gaussian estimation by accounting for the third and fourth statistical moments $m_{3}$ and $m_{4}$.
\end{itemize}

\begin{figure}[tbhp]
\centering
\includegraphics[width=1\linewidth]{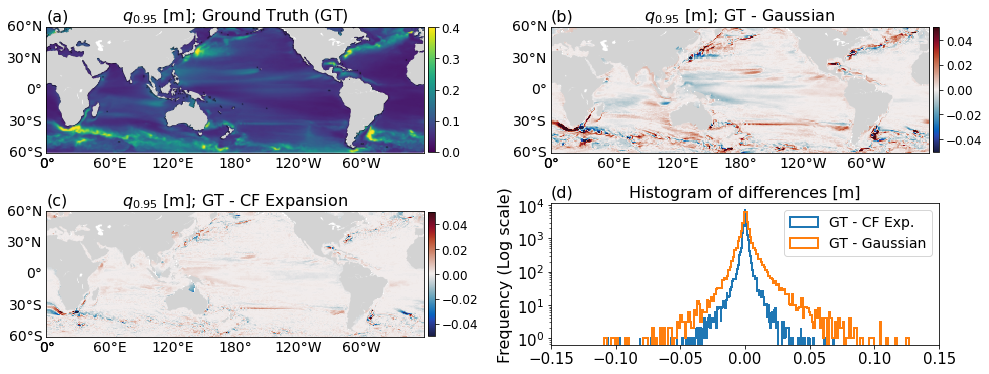}
\caption{A test for the Cornish-Fisher expansion in the case of dynamic sea level. Panel (a): we estimate the quantile 0.95 ($q_{0.95}$) for each sea level time series $\eta^{\mbox{\tiny dyn}}(x,y,t)$ with the Python, NumPy quantile function. We refer to this estimation as the ``Ground Truth'' or ``GT''. Panel (b): we estimate $q_{0.95}$ under the assumption of Gaussian statistics. In this case, the mean $m_{1}$ and variance $m_{2}$ of each time series $\eta^{\mbox{\tiny dyn}}(x,y,t)$ is enough to compute any quantile. Panel (c): we correct the estimation of $q_{0.95}$ in panel (b) through the Cornish-Fisher expansion. Panel (d): we consider the histogram of differences between the ``Ground Truth'' (panel (a)) and the Gaussian and Cornish-Fisher estimations shown in panel (b,c). Note how the differences are greatly reduced using the CF expansion relative to the Gaussian expansion}
\label{fig:cf_expansion_sea_level}
\end{figure}

Results are shown in Figure \ref{fig:cf_expansion_sea_level}. In the case of quantiles estimated through Cornish-Fisher we always obtain closer values to the ``Ground Truth'' in respect to the Gaussian estimation. This result is clearly shown in the histograms of differences between the ``Ground Truth'' and the Gaussian and Cornish-Fisher estimations, respectively in orange and blue (Figure \ref{fig:cf_expansion_sea_level}(d)).\\

\clearpage

\section{Tide gauges: p-values and False discovery rate} \label{app:p-values-tide-gaues}

\begin{figure}[tbhp]
\centering
\includegraphics[width=1\linewidth]{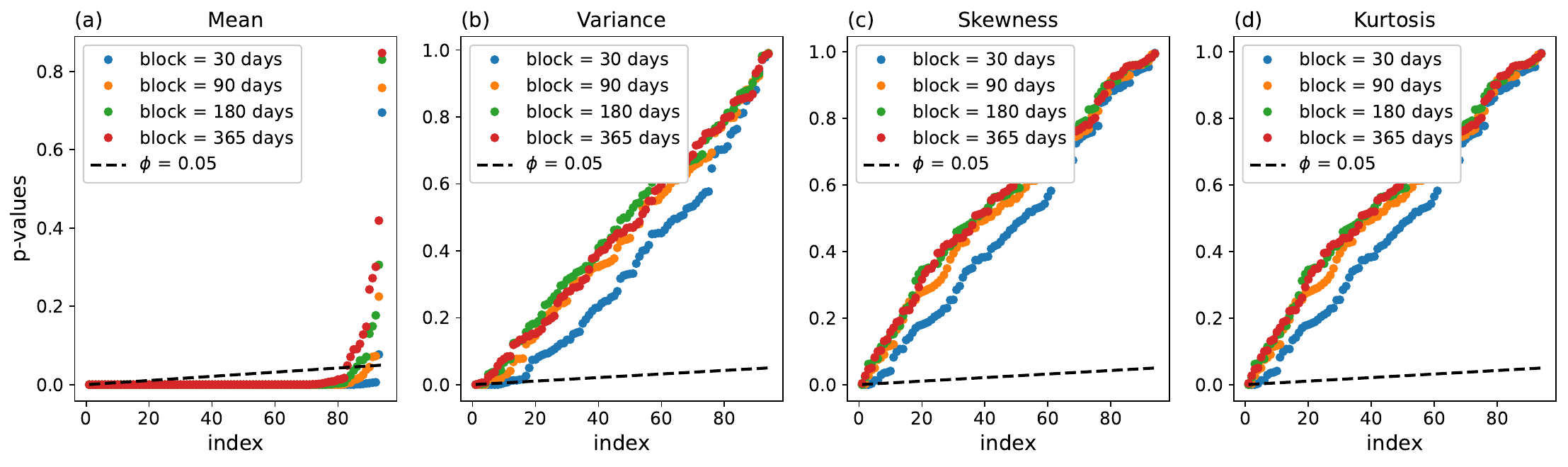}
\caption{p-values for tide gauges data in the 1970-2017 period. Panels (a-d): sorted p-values for changes in distributional mean, variance, skewness and kurtosis respectively. p-values have been computed from the block-bootstrapped distribution. These plots investigate the robustness of the p-value computation under blocks of 30, 90, 180 and 365 days are reported. Convergence is achieved starting from 90 days (block chosen in the analysis). The False Discovery Rate $\phi = 0.05$ is also reported}
\label{fig:p-values_Tide_Gauges_1970_2017}
\end{figure}

In this work, p-values are computed from the block-bootstrapped distribution (see Section \ref{sec:significance_test}). Here we analyze the robustness of block size when computing p-values and present results in Figure \ref{fig:p-values_Tide_Gauges_1970_2017}. Results are robust in the case of blocks of one season (90 days), six months and one year while diverging for smaller sizes such as 30 days. The results shown in the main paper using 1 season are then robust and this choice is adopted throughout the paper.

\clearpage

\section{Tide gauge data at Balboa, Panama} \label{app:balboa_record}

\begin{figure}[tbhp]
\centering
\includegraphics[width=0.8\linewidth]{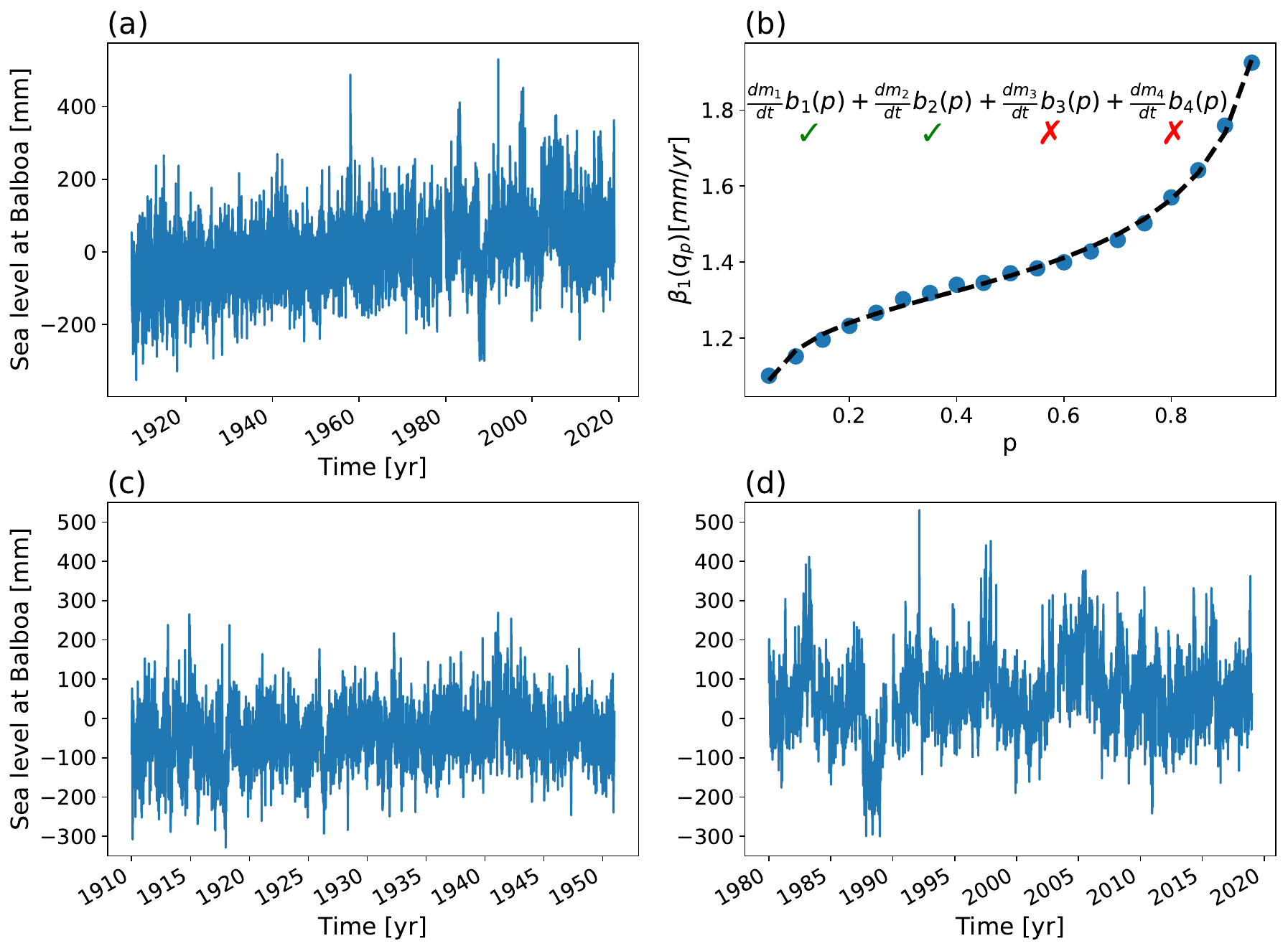}
\caption{Analysis of the Balboa, Panama tide gauge from 1907-06-20 to 2018-12-31. Panel (a): recorded sea level at Balboa, Panama. Panel (b): linear quantile slopes for the time series shown in panel (a). The slopes $\beta_{1}(q_{p})$ are computed for $p \in [0.05, 0.95]$ every $dp = 0.05$ and indicated as blue dots. The black, dashed line indicates the projection onto polynomials as defined in Eq. \eqref{eq:projection}. Green ``check'' marks indicate statistically significant ($95\%$ confidence level) changes in moments; red "crosses" indicate non-significant changes. Panel (c,d): time series in the first and last 40 years of recorded sea level}
\label{fig:balboa_panama}
\end{figure}

The only two long-record tide gauges with changes in higher order moments were both found in Panama. In Figure \ref{fig:balboa_panama}, we show one of the two tide gauges for which we identified a statistical significant change in variance.\\

The full time series is shown in Figure \ref{fig:balboa_panama}(a) and the quantile slopes together with the projection onto polynomials is shown in in Figure \ref{fig:balboa_panama}(b). The first and last 40 years of the time series are plotted in Figure \ref{fig:balboa_panama}(c,d). Figure \ref{fig:balboa_panama}(d) clearly show larger variance in respect to the first 40 years. At least in part, these changes can be explained by large sea level oscillations forced by the El Ni\~no Southern Oscillation (ENSO). Large, positive sea level anomalies can be clearly identified across the years 1982/1983 and 1997/1998 where two large El Ni\~no were recorded \citep{historicalENSO}. Very large, persistent negative sea level anomalies can be seen across the years 1988/1989 in correspondence of a strong La Ni\~na event.\\

\begin{figure}[tbhp]
\centering
\includegraphics[width=1\linewidth]{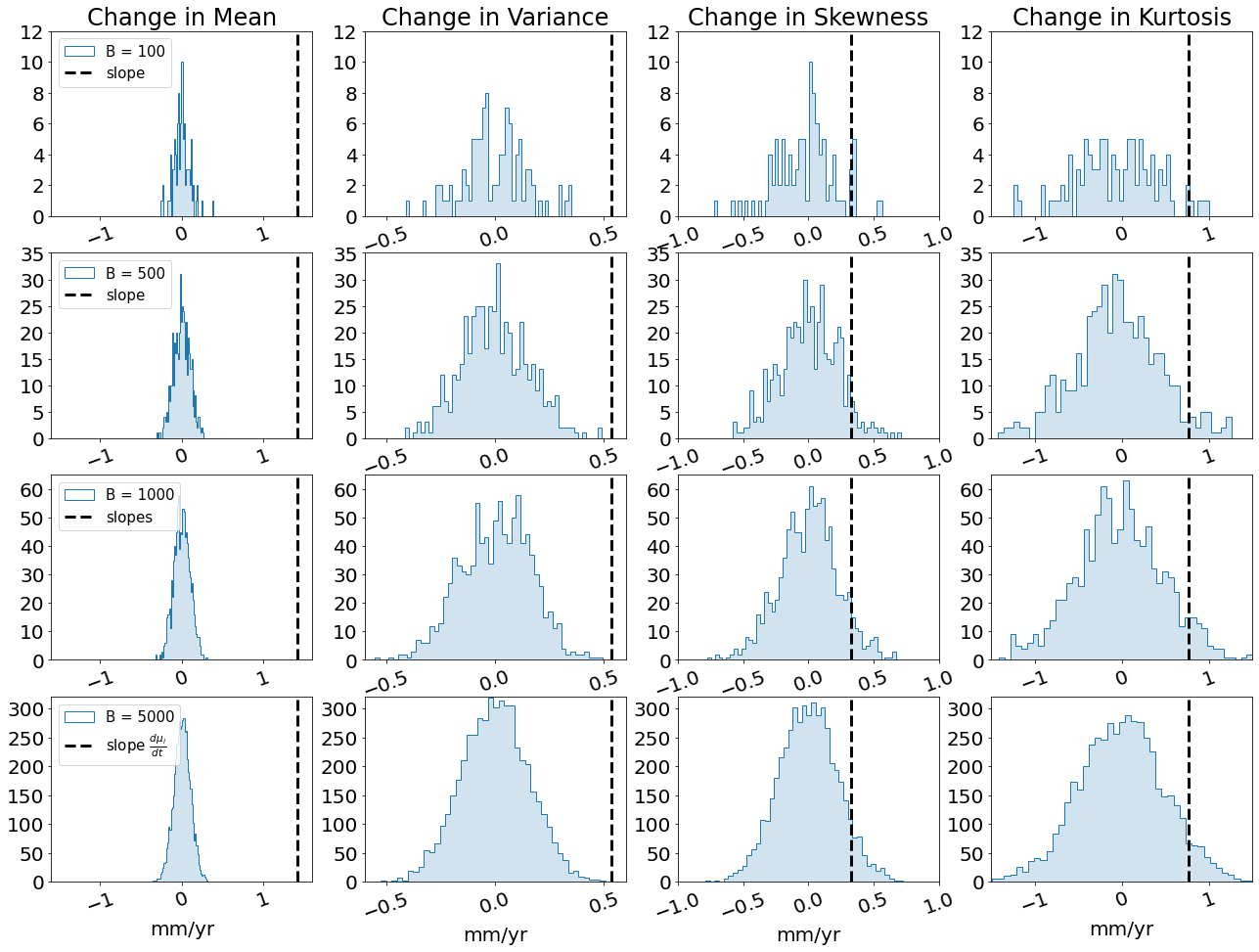}
\caption{Analysis of the Balboa, Panama tide gauge from 1907-06-20 to 2018-12-31. Here we show the bootstrapped distribution to infer statistical significance in changes in the mean, variance, skewness and kurtosis (number of bins is set to 50). The dashed black lines indicate the slopes in moments. We focus on the dependence of our analysis to the sample size $B$ used for to infer the \textit{null} distribution. Here $B$ ranges from $B = 100$ (first row) to $B = 5000$ (fourth row).  In all examples, the mean and variance show significant changes at the $95\%$ level while changes in skewness and kurtosis are found to be not significant. This means that bootstrapping with sample sizes as low as $B = 100$ can still give meaningful results. In all our analysis in this paper we used $B = 1000$}
\label{fig:bootstrap_tests}
\end{figure}

In Figure \ref{fig:bootstrap_tests} we provide a justification for the choice of sample size $B$ to infer the bootstrapped distribution. The bootstrapped distribution is then used to estimate the statistical significance in changes in mean, variance, skewness and kurtosis (see Section \ref{sec:methodology}). Here we show that even with sample sizes as small as $B = 100$ we obtain the same results in terms of statistical significance as with $B = 5000$ when looking at $95\%$ confidence level. Generally, larger sample sizes are always preferred and in this paper we adopt $B = 1000$. Nonetheless, when focusing on many time series it could be useful considering smaller $B$ for faster computation.\\

\clearpage

\section{Time series in the GFDL-CM4 Mediterranean sea} \label{app:Med-sea}

\begin{figure}[tbhp]
\centering
\includegraphics[width=1\linewidth]{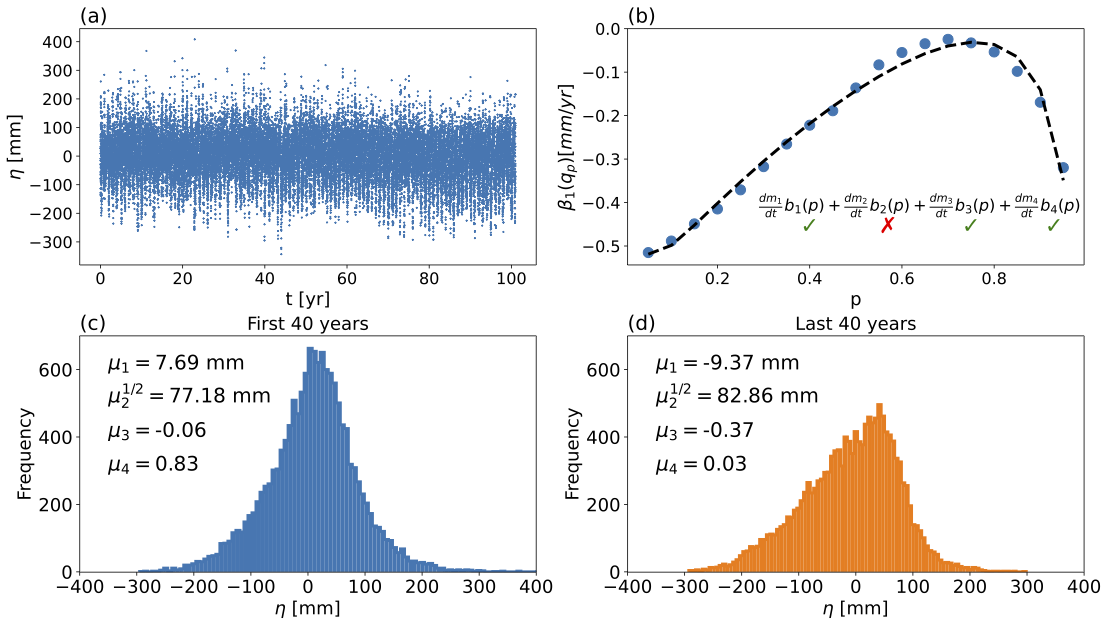}
\caption{Panel (a): time series showing the largest changes in kurtosis in the Mediterranean sea for the GFDL-CM4 model 1pctCO2 run. Panel (b): linear quantile slopes for the time series shown in panel (a). The slopes $\beta_{1}(q_{p})$ are computed for $p \in [0.05, 0.95]$ every $dp = 0.05$ and indicated as blue dots. The black, dashed line indicates the projection onto polynomials as defined in Eq. \eqref{eq:projection}. Green ``check'' marks indicate statistically significant ($95\%$ confidence level) changes in moments; red "crosses" indicate non-significant changes. Panel (c,d): histograms for the first and last 40 years of data of the time series shown in panel (a)}
\label{fig:mediterrenean_sea}
\end{figure}

The 1pctCO2 simulation shows many changes in higher order moments. Numerous changes in the third and fourth moments are found in the Mediterranean sea. Here, we extract the time series in the (GFDL-CM4) Mediterranean with largest changes in kurtosis (fourth moment). We then plot the probability distributions in the first and last 40 years of the simulation showing the large changes in shapes of distributions.  This result is shown in Figure \ref{fig:mediterrenean_sea}. 

\clearpage

\subsubsection*{Code availability}
Codes and materials are available at \url{https://zenodo.org/badge/latestdoi/623102086}. 

\subsubsection*{Acknowledgments}

F.F. acknowledges discussions on the topic with Elizabeth Yankovsky, Aaron Match and Simone Contu and Ryan Sh\`iji\'e D\`u. F.F. acknowledges a correspondence with Blaise Melly with helpful suggestions on fast algorithms for the quantile regression step. We also thank John Krasting and Jan-Erik Tesdal for useful comments on an early draft.  Additionally, this work was supported through the NYU IT High Performance Computing resources, services, and staff expertise. 

\subsubsection*{Funding statement}
The work was supported through NOAA grant NOAA-OAR-CPO-2019-2005530 and by the VoLo foundation. This research was supported by Schmidt Futures, a philanthropic initiative founded by Eric and Wendy Schmidt, as part of its Virtual Earth System Research Institute (VESRI). 


\bibliographystyle{unsrtnat}
\bibliography{references}

\end{document}